\documentclass[%
 reprint,
 superscriptaddress,
 amsmath,amssymb,
 aps,
prb,
floatfix,
]{revtex4-1}

\usepackage{graphicx}
\usepackage{dcolumn}
\usepackage{bm}
\usepackage[normalem]{ulem}
\usepackage{hyperref}
\hypersetup{colorlinks = true, citecolor = blue, breaklinks = true}
\usepackage{textcomp}
\usepackage{multirow}
\usepackage{xfrac}
\newcommand{\angstrom}{\mbox{\normalfont\AA}}
\usepackage[resetlabels]{multibib}
\newcites{SI}{References}

\begin{document}

\title{Self-consistent site-dependent DFT+$U$ study of stoichiometric and defective SrMnO$_3$}

\affiliation{%
Department of Chemistry and Biochemistry, University of Bern, Freiestrasse 3, CH-3012 Bern, Switzerland 
}%
\affiliation{%
Theory and Simulation of Materials (THEOS), Ecole Polytechnique F\'ed\'erale de Lausanne, CH-1015 Lausanne, Switzerland
}%
\affiliation{%
National Centre for Computational Design and Discovery of Novel Materials (MARVEL), Switzerland
}%

\author{Chiara Ricca}
\affiliation{%
Department of Chemistry and Biochemistry, University of Bern, Freiestrasse 3, CH-3012 Bern, Switzerland 
}%
\affiliation{%
National Centre for Computational Design and Discovery of Novel Materials (MARVEL), Switzerland
}%

\author{Iurii Timrov}
\author{Matteo Cococcioni}
\author{Nicola Marzari}
\affiliation{%
Theory and Simulation of Materials (THEOS), Ecole Polytechnique F\'ed\'erale de Lausanne, CH-1015 Lausanne, Switzerland
}%
\affiliation{%
National Centre for Computational Design and Discovery of Novel Materials (MARVEL), Switzerland
}%

\author{Ulrich Aschauer}
\email{ulrich.aschauer@dcb.unibe.ch}
\affiliation{%
Department of Chemistry and Biochemistry, University of Bern, Freiestrasse 3, CH-3012 Bern, Switzerland 
}%
\affiliation{%
National Centre for Computational Design and Discovery of Novel Materials (MARVEL), Switzerland
}%

\date{\today}

\begin{abstract}
We propose a self-consistent site-dependent Hubbard $U$ approach for DFT+$U$ calculations of defects in complex transition-metal oxides, using Hubbard parameters computed via linear-response theory. The formation of a defect locally perturbs the chemical environment of Hubbard sites in its vicinity, resulting in different Hubbard $U$ parameters for different sites. Using oxygen vacancies in SrMnO$_3$ as a model system, we investigate the dependence of $U$ on the chemical environment and study its influence on the structural, electronic, and magnetic properties of defective bulk and strained thin-film structures. Our results show that a self-consistent $U$ improves the description of stoichiometric bulk SrMnO$_3$ with respect to GGA or GGA+$U$ calculations using an empirical $U$. For defective systems, $U$ changes as a function of the distance of the Hubbard site from the defect, its oxidation state and the magnetic phase of the bulk structure. Taking into account this dependence, in turn, affects the computed defect formation energies and the predicted strain- and/or defect-induced magnetic phase transitions, especially when occupied localized states appear in the band gap of the material upon defect creation.
\end{abstract}

\maketitle

\section{\label{sec:intro}Introduction}

Defect chemistry has recently emerged as a parameter for the design and discovery of novel functional materials \cite{marthinsen2016coupling,griffin2017defect}. Strongly correlated oxides, in particular ABO$_3$ perovskites, show a wide spectrum of technologically relevant functional properties that can often be tuned via strain, imposed for example by lattice matching with the substrate during coherent epitaxial growth of thin films \cite{Meng201707817, haeni2004}. Defects can affect the ground state properties of these materials, promote phase transitions or enable entirely new functionalities \cite{fuchigami2009, tuller2011, kalinin2012, kalinin2013functional, chandrasekaran2013, adler2004, bivskup2014, bhattacharya2014magnetic, becher2015strain, rojac2017domain}. Besides altering the functional properties of the material, strain was shown to also influence the defect chemistry, tensile strain generally favoring the formation of anion vacancies and compressive strain promoting cation vacancies \cite{aschauer2013strain, gan2014anisotropic, aidhy2014strain, choi2015assessment, aschauer2015effect}. This suggests the existence of a rich phase diagram, in which functional properties such as ferroelectricity and magnetism and the defect chemistry are altered by strain, while also coupling or competing with each other in determining a material's properties \cite{kalinin2012, kalinin2013functional, marthinsen2016coupling, becher2015strain, aschauer2016interplay, chandrasena2017strain}.

\begin{figure}
 \centering
 \includegraphics[width=0.9\columnwidth]{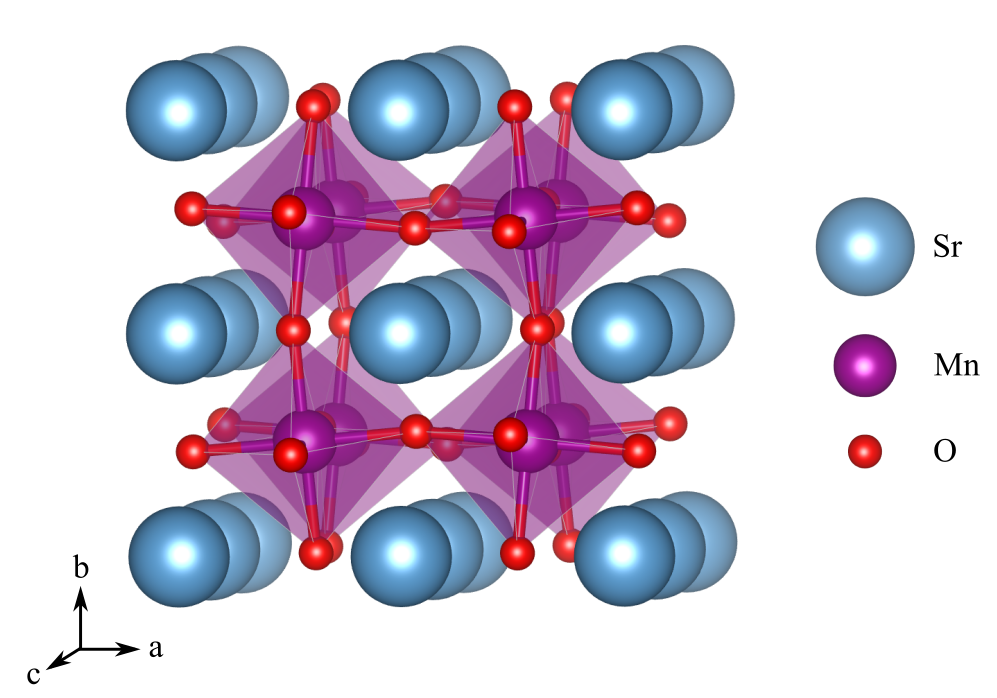}
 \caption{(2$\times$2$\times$2) \textit{Pnma} supercell of stoichiometric SrMnO$_3$. Apart from the G-type antiferromagnetic order, all Mn atoms are equivalent.}
\label{fig:SMO_structure_bulk}
\end{figure}

SrMnO$_3$ (SMO) is an interesting example for defect-enabled functionality, since oxygen-vacancy formation energies in this material result from a complex interplay of strain, magnetic order and polar distortions \cite{marthinsen2016coupling}. While the low-temperature ground state of SMO is the 4H hexagonal phase \cite{Sondena2006}, the perovskite structure is stable at high temperatures \cite{Nielsen2014} and can be stabilized at low temperature in thin films \cite{Kobayashi:2010by}. The perovskite phase (space group $Pnma$, see Fig.~\ref{fig:SMO_structure_bulk}) has a G-type antiferromagnetic (AFM)~\cite{chmaissem2001relationship} order of the Mn magnetic moments and is close to the ideal cubic structure, with small octahedral rotations found computationally~\cite{lee2010epitaxial} but not yet observed experimentally.

Strain-engineering the properties of SMO has also been a subject of interest. It was predicted  from theory that tensile strain can induce ferroelectricity~\cite{lee2010epitaxial, becher2015strain} while simultaneously increasing the oxygen-vacancy concentration in SMO thin films~\cite{becher2015strain, marthinsen2016coupling}. These strain-induced defects can couple with strain-induced ferroelectric domain walls and form barriers for electrical conductivity~\cite{becher2015strain}. Strain-dependent magnetic properties of SMO thin films are still an open question. Some theoretical and experimental results~\cite{lee2010epitaxial, becher2015strain, maurel2015nature} have found SMO thin films to be AFM, while more recent experimental studies~\cite{wang2016oxygen, bai2017structural} report that SMO thin films have ferromagnetic (FM) properties. Ferromagnetism was explained to emerge from strain-induced changes of the Mn-O-Mn bond angles and the Mn$^{3+}$--Mn$^{4+}$ double exchange coupling induced by the presence of oxygen vacancies.

This last observation is in agreement with the theoretical prediction that a 4.2\% oxygen-vacancy concentration in the unstrained perovskite SMO structure can stabilize the FM over the stoichiometric AFM ground state \cite{marthinsen2016coupling}. That work was performed using density-functional theory (DFT) with a Hubbard $U$ correction~\cite{anisimov1991band} whose value was adjusted to reproduce the electronic and magnetic properties of stoichiometric manganites \cite{Hong2012, aschauer2013strain}. This may lead to potential shortcomings in the description of the electronic structure of defective SMO. While approaches such as DFT+$U$ are required to correct self-interaction in complex oxides with localized $d$ electrons, it is not clear if the absence of a band gap predicted by DFT for FM SMO~\cite{marthinsen2016coupling} is a real indicator of metallicity or a consequence of the fact that neither approximate nor exact DFT can predict band gaps. More importantly, a further potential deficiency is related to the description of multiple oxidation states present in defective SMO. The formation of oxygen vacancies is generally charge compensated by a reduction of the oxidation state (OS) of some Mn$^{4+}$ to Mn$^{3+}$; transition metals with different OS may not be properly described by the same $U$ \cite{Cococcioni2005, Franchini2007, Santana2014, Lu2014a}.

Taking into account local structural and chemical effects for each transition metal site in the oxide may thus be crucial to achieve an accurate description of defective SMO. In this work, we study the defect chemistry and magnetic properties of SMO bulk and strained thin films using a theoretical approach that takes this aspect intrinsically into account by performing DFT+$U$ calculations with self-consistent $U$ values computed from first principles for each inequivalent Hubbard site in the structure. This self-consistent site-dependent (SC-SD) DFT+$U$ approach will be described in detail below. Due to the lack of experimental reference data for the electronic properties and the oxygen-vacancy formation energies for SMO, we will apply DFT+$U_\mathrm{SC-SD}$ to both stoichiometric and defective SrMnO$_3$ and compare our results with standard DFT or DFT+$U$ (the latter with empirically derived Hubbard parameters) to investigate the impact of DFT+$U_\mathrm{SC-SD}$ on the predictions for the structure, phase stability, electronic, and magnetic properties, and on the defect energetics.

\section{{\label{sec:dftmethods}Self-interaction corrections in DFT}}

DFT is a powerful tool for materials design. However, standard DFT approaches based on the local-density (LDA) or generalized-gradient approximations (GGA) of the exchange-correlation functional are known to provide inaccurate predictions when treating strongly correlated systems, such as transition-metal oxides with localized \textit{d} or \textit{f} states. The failure of these functionals is mainly due to the self-interaction error that is  present in approximate DFT, resulting often in electron over-delocalization, while leading to electronic levels and  band gaps (neither of which are meant to be described exactly by DFT) further away from estimates obtained with more modern functionals~\cite{cohen2008insights}. These quantities are, however, crucial for the description of a defect's electronic structure and LDA and GGA defect calculations in strongly self-interacting and typically also correlated materials are hence generally not that accurate. More advanced functionals such as hybrid functionals (\textit{e.g.} HSE06)~\cite{zhao2005exchange, heyd2004efficient} or  the DFT+$U$ method \cite{anisimov1991band} can greatly improve upon standard LDA/GGA.

Hybrid functionals incorporate a fraction of exact (Fock) exchange into standard DFT functionals, improving the agreement with experiment due to the cancellation between the GGA overestimation and the Hartree-Fock underestimation of delocalization~\cite{ciofini2004}. Unfortunately, due to the non-local nature of the Fock exchange, calculations with hybrid functionals are computationally quite expensive for practical applications in solid-state physics, especially with plane-wave codes and for defect calculations, which require  large supercells. It is also known that including exact exchange can, in some cases, decrease  accuracy, especially for metals where the unphysical logarithmic singularity of band energies at the Fermi level results in spurious charge and spin densities \cite{paier2007, mosey2008rotationally, ong2011comparison}. Moreover, the fraction of Fock exchange to use is a material-dependent parameter, which in practice is often determined empirically by fitting to experimental data; however different experimental properties  result in different fractions of Fock exchange \cite{ciofini2004, Franchini2007, Getsoian2013, Franchini2014}. Not only do lattice parameters and band gaps \cite{Zhang2005hybrid, Franchini2014, onishi2008hybrid} show a different dependence on the fraction of Fock exchange, but in the case of defects, the position of the defect levels and consequently the computed formation energies are also affected~\cite{Zhang2005hybrid}. Furthermore, the description of multivalent oxides or defective systems containing metal ions with different chemical environments or oxidation states may not be straightforward, since the fraction of Fock exchange in a standard hybrid functional is a global quantity, while it was observed that the same metal in different oxidation states requires different fractions of Fock exchange \cite{Franchini2007}.

Hubbard corrected functionals~\cite{anisimov1991band, anisimov1997first,Dudarev1998} are a popular alternative to study these systems, due to the simplicity of their formulation, their modest computational costs (only slightly larger than LDA or GGA) and an intuitive physical picture. In this approach, strongly localized \textit{d} and \textit{f} electrons are described in analogy to the Hubbard model, while the rest of the valence electrons are treated at the LDA or GGA level: a simple corrective term is added to the DFT energy functional, depending on the effective on-site Coulomb interaction parameter (the so-called Hubbard $U$) that aims to improve interactions between strongly localized electrons.

A key issue in DFT+$U$ is the choice of the Hubbard $U$ parameter, which describes the strength of these interactions and which is \textit{a priori} unknown. $U$ is often considered as an empirical parameter that can be obtained by fitting to experimental quantities of interest, such as band gaps, charge localization, kinetic barriers or thermodynamics quantities \cite{Andersson2007, morgan2007dftu, dasilva2007hybrid, laubach2007, scanlon2009acceptor}. However, not only does this procedure limit the application of DFT+$U$ to systems for which reliable experimental information is available, but the choice of an empirical $U$ is not unique, as fitting different experimental properties leads to different $U$ values \cite{Getsoian2013, Capdevila-Cortada2016}. For the case of defects, it was shown that oxygen-vacancy formation energies strongly depend on the chosen $U$ \cite{fabris2005taming, Fabris2005, morgan2007dftu, Andersson2007, Lutfalla2011, Curnan2015} and it was concluded that fitting to band gaps or structural parameters will not necessarily provide defect energetics that agree with experiments \footnotetext{On the other hand, it should be noted that the experimental determination of accurate defect formation energies is a complex task with inherent error bars.} \cite{Lutfalla2011,Wang2006,Note1}. Furthermore, $U$  depends on the local chemical environment of the Hubbard site, \textit{i.e.} its oxidation state \cite{Cococcioni2005, Franchini2007} or spin state \cite{Kulik2006, Scherlis2007, mosey2008rotationally, Kulik2010}. However, in conventional DFT+$U$ calculations a ``global'' $U$ value is applied to all transition-metal atoms, similarly to hybrids where a ``global'' fraction of the Fock exchange is applied (albeit to all atoms). Some authors pointed out that a ``global'' $U$ for Hubbard sites with different chemical environments may not be appropriate \cite{Cococcioni2005, Franchini2007, Santana2014, Lu2014a}. This is clearly the case for defective systems, where defects can induce localized or delocalized states on surrounding transition-metal atoms. A site-dependent DFT+$U$ approach \cite{Franchini2007, Lu2014a} based on empirical $U$ values is, however, extremely difficult to implement as  experimental data are usually insufficient for this task. Last but not least, a DFT+$U$ formulation with empirical $U$ values cannot be considered a fully \textit{ab initio} approach and it cannot be applied to systems for which no experimental data is available, hence it cannot be used as a predictive tool for new materials.

An alternative strategy is to compute $U$ from first principles without relying on experimental data. Indeed, $U$ can be seen not as an empirical fitting parameter, but as an intrinsic response property of the system, as argued by Cococcioni and de Gironcoli in Ref.~\onlinecite{Cococcioni2005}, who devised a constrained DFT (cDFT) approach using linear-response (LR) theory to compute $U$ from first principles. This method derives Hubbard parameters from the response of the Hubbard sites' electronic occupations to a small perturbation of the  potential acting on the Hubbard manifold of the $d$ or $f$ electrons. While conceptually very simple, LR-cDFT is quite cumbersome in practice, because it requires supercells and is based on finite differences, with rather involved post-processing. Recently, Timrov \textit{et al.}~\cite{Timrov2018} developed a novel formulation for the calculation of $U$ using density-functional perturbation theory (DFPT), which remains equivalent by construction to LR-cDFT. The DFPT approach has several advantages over LR-cDFT since neither supercells nor finite differences are needed -- instead, primitive unit cells with monochromatic perturbations are used, which considerably speeds up the calculation of $U$, while automating the post-processing of the results. Despite the fact that defect calculations require supercells and we loose the advantage of being able to use primitive unit cells, DFPT with \textbf{q}-meshes is still more affordable and offers higher precision and better convergence of $U$ compared to LR-cDFT~\cite{Timrov2018}.

It is important to note that the actual value of Hubbard $U$ computed from first-principles can vary by a fraction of an eV or up to several eV depending on the choice of localized functions used to construct the Hubbard manifold (\textit{e.g.} atomic, orthogonalized atomic, Wannier, etc.), the exchange-correlation functional used, types of pseudopotentials, etc.~\cite{pickett1998, Fabris2005, Wang2006, Kulik2008,Kulik2010}. This means that $U$ should not be seen as a transferable, universal parameter for a certain chemical element or chemical composition. For consistency, DFT+$U$ calculations must be performed using the same computational setup that was used for the calculation of $U$. Computing $U$ from first principles not only has the advantage of being predictive for novel materials, but it also allows to take into account local structural and chemical effects through the calculation of site-dependent $U$ parameters for different Hubbard sites in the structure. A key point in this approach is that $U$ can be seen as a functional of the electron density and consequently it must be computed self-consistently, taking into account also structural effects \cite{Zhou2004, Hsu2009, shishkin2016self}. Here, by ``self-consistency'' we mean that $U$ and the geometry are recomputed in an iterative way starting from the DFT level until both converge within given thresholds. This is required because the LR-DFPT $U$ depends on the electronic ground state, the atomic positions and cell vectors, which implies that forces and stresses have additional components related to the derivative of $U$ with respect to positions or strain. We observed major changes in geometry when the applied $U$ resulted in a different electronic ground state (\textit{e.g.} when going from the DFT ground state to the first DFT+$U$ solution) but expect smaller changes during the self-consistent optimization of both the structure and $U$ as derivatives of $U$ with respect to positions or strain become smaller.  We note here that for a true variational solution, the Hubbard parameters should be calculated concurrently within the Kohn-Sham self-consistent cycle, which is currently not available. The present self-consistent scheme ensures internal consistency of the results and therefore currently represents the best possible compromise. Computing in this way self-consistent structure-dependent $U$ ($U_\mathrm{SC}$) for the bulk, and self-consistent site-dependent ($U_\mathrm{SC-SD}$) for the transition-metal atoms around a defect, it is possible to take into account changes of $U$ due to structural relaxations and excess charge localization in defective structures. This approach provides a more accurate description than standard DFT and possibly hybrid functionals at a computational cost that is significantly lower than for the latter. We note here that site dependence has also been implemented in some recently developed hybrid functionals~\cite{jaramillo2003,kaupp2006,kaupp2007,kaupp2007b,kaupp2007c,shimazaki2015theoretical}.

The SC-SD DFT+$U$ approach is similar in spirit to the pseudohybrid Hubbard DFT functional ACBN0~\cite{agapito2015}, in which DFT+$U$ calculations are performed using Hubbard parameters evaluated self-consistently for every atom via direct computation of on-site Coulomb and exchange energies. In that approach, as in SC-SD DFT+$U$, the Hubbard correction is a natural function of the electron density and of the chemical environment, and different Hubbard $U$ parameters are obtained for inequivalent sites~\cite{lee2017systematic}. It should be noted that ACBN0 $U$ values are adjusted to reproduce Hartree-Fock (HF) electron-electron interactions, while the present scheme includes correlations at the DFT+$U$ level.

Last, we conclude reminding that DFT is a theory of total energies so fitting a functional to reproduce electronic states (\textit{e.g.} bad gaps) is debatable; in this sense, approaches where $U$ is chosen based on energetic configurations seems more appropriate.

\section{Methods}
\label{sec:compdetails}

DFT+$U$ calculations were performed using the simplified formulation of Dudarev {\it et al.}~\cite{Dudarev1998} as implemented in the {\sc{Quantum ESPRESSO}}  distribution~\cite{giannozzi2009quantum,Giannozzi2017}. We used GGA for the exchange-correlation functional with the PBEsol parameterization~\cite{perdew2008pbesol}, and ultrasoft pseudopotentials \cite{vanderbilt1990soft} with Sr($4s$, $4p$, $5s$), Mn($3p$, $4s$, $3d$), and O($2s$, $2p$) valence states \footnote{Ultrasoft pseudopotentials from the PSLibrary were taken from \url{www.materialscloud.org}: Sr.pbesol-spn-rrkjus\textunderscore psl.1.0.0.UPF, Mn.pbesol-spn-rrkjus\textunderscore psl.0.3.1.UPF, and O.pbesol-n-rrkjus\textunderscore psl.1.0.0.UPF}. A kinetic-energy cutoff of 70 Ry for wave functions and 840 Ry for spin-charge density and potentials were used. A Gaussian smearing with a broadening parameter of 0.01 Ry was used in all cases, including plotting the density of states (DOS).

In order to reproduce the G-AFM magnetic order, we studied SMO using a 40-atom $2\times2\times2$ supercell of the 5-atom primitive cubic cell, using a shifted $6\times6\times6$ Monkhorst-Pack \cite{monkhorst1976special} \textbf{k}-point grid to sample the Brillouin zone. A finer $8\times8\times8$ grid was used when plotting the DOS. We considered G-type AFM and FM phases, both in bulk and thin film geometries. For stoichiometric bulk calculations, both lattice parameters and atomic positions were relaxed, while thin film geometries with biaxial epitaxial strain in the $ac$ plane imposed by a cubic substrate were computed following the procedure described in Ref. \onlinecite{Rondinelli:2011jk}. Namely, lattice parameters \textit{a} and \textit{c} were adjusted to the desired strain and kept fixed at equal length and at 90$^{\circ}$ to each other, while the out-of-plane \textit{b} axis and the atomic positions were allowed to relax. Defects were created by removing one oxygen atom from the 40-atom supercell (vacancy concentration 4.2\%). Neutral (V$_\textrm{O}^{\bullet \bullet}$), singly (V$_\textrm{O}^{\bullet}$), and doubly (V$_\textrm{O}^\textrm{X}$) positively charged oxygen-vacancy defects were studied by adjusting the number of electrons. For charged defects a background charge ensuring neutrality of the unit cell is present, as implied in calculations under periodic-boundary conditions to avoid divergences in the electrostatic potential. For defect calculations, atomic positions were optimized while keeping the lattice vectors fixed at optimized values for stoichiometric bulk or strained SMO. In all cases, convergence thresholds of 5$\times$ 10$^{-2}$ eV/\angstrom{} for the atomic forces and 1.4 $\times$ 10$^{-5}$ eV for the energy were used.

The Hubbard correction~\cite{anisimov1991band, anisimov1997first, Dudarev1998} was applied to Mn-3\textit{d} states with an empirical $U$ ($U_{E}$) of 3.0 eV \cite{jung1997determination,Hong2012} and by using self-consistent (site-dependent) $U$ parameters computed using DFPT ~\cite{Timrov2018}, as implemented in {\sc{Quantum ESPRESSO}}~\cite{giannozzi2009quantum,Giannozzi2017}. 
Within linear-response theory, Hubbard parameters are computed as the difference between bare and screened inverse susceptibilities~\cite{Cococcioni2005}: 
\begin{equation}
U^I = \left(\chi_0^{-1} - \chi^{-1}\right)_{II} \,,
\label{eq:Ucalc}
\end{equation}
where $I$ is an atomic site index. The susceptibilities $\chi_0$ and $\chi$ measure the response of atomic occupations to shifts in the potential acting on all the Hubbard states of single atoms and are defined as $\chi_{IJ} = \sum_{m\sigma} \left(dn^{I\sigma}_{mm} / d\alpha^J\right)$, where $n^{I\sigma}_{mm'}$ are atomic occupation matrices, $\alpha^J$ is the strength of the perturbation on the site $J$, $m$ and $m'$ are magnetic quantum numbers associated with a specific angular momentum, and $\sigma$ is the spin index. $\chi$ is evaluated at self-consistency (of the linear-response Kohn-Sham calculation), while $\chi_0$, its non-interacting counterpart, is computed before the self-consistent re-adjustment of the Hartree and exchange-correlation potentials. The basic idea of the DFPT implementation is to recast the entries of the susceptibility matrices into sums of monochromatic contributions over the Brillouin zone~\cite{Timrov2018}:
\begin{equation}
\frac{dn^{I\sigma}_{mm'}}{d\alpha^J} = \frac{1}{N_\mathbf{q}}\sum_\mathbf{q}^{N_\mathbf{q}} e^{i\mathbf{q}\cdot(\mathbf{R}_l - \mathbf{R}_{l'})}\Delta_\mathbf{q}^{s'} n^{s \,\sigma}_{mm'} \,,
\label{eq:dnq}
\end{equation}
where $I\equiv(l,s)$ and $J\equiv(l',s')$; $l$ and $l'$ label unit cells, $s$ and $s'$ represent atoms in the unit cell, and $\mathbf{R}_l$ and $\mathbf{R}_{l'}$ are Bravais lattice vectors.  $\Delta_\mathbf{q}^{s'} n^{s \,\sigma}_{mm'}$ represent the lattice-periodic response of atomic occupations to monochromatic perturbations of wavevector $\mathbf{q}$.

In periodic systems without defects, this approach allows to use primitive cells rather than computationally expensive supercells~\cite{Timrov2018}. When supercells are needed, for example for systems with defects as in this work, the DFPT calculation of the Hubbard parameters, using coarser $\mathbf{q}$-point grids than for primitive cells, still offer a higher level of accuracy, automation, and a better scaling than the linear response approach of Ref.~\cite{Cococcioni2005}. In the present study we found that a zone center ($\mathbf{q}=0$) perturbation was enough to converge the final $U$ to within 0.001 eV.

Self-consistent $U$ parameters are calculated using an iterative procedure that starts with a structure optimization at the GGA level ($U$=0 eV). $U$ is then computed starting from this GGA ground state using DFPT. Subsequently, a GGA+$U$ geometry optimization is performed with the computed $U$, followed by another DFPT calculation to determine a new $U$ value for this optimized structure, which will be used in the next iteration. The procedure is stopped when the structure remains the same and convergence of the Hubbard parameter is achieved to within a desired precision (in this study a convergence threshold of 0.01 eV for the self consistency of $U$ was used). For stoichiometric SMO calculations, all Mn sites are crystallographically and chemically equivalent and can thus be described with a single global $U$ value ($U_\mathrm{SC}$), computed with the procedure described above. For defective systems, instead, a self-consistent {\it site-dependent} calculation of $U$ is performed: $U$ is computed based on the distance of the Hubbard Mn site from the defect and on its chemical environment ($U_\mathrm{SC-SD}$). In all cases, atomic orbitals were used to construct occupation matrices and projectors in the DFT+$U$ scheme.

The strain-dependent formation energy of an oxygen vacancy ($\textrm{V}_\textrm{O}$) in a charge state $\textrm{q}$ ($\textrm{E}_{\textrm{f},\textrm{V}_\textrm{O}^\textrm{q}}$) was calculated as described in Ref.~\onlinecite{freysoldt2014first}:
\begin{multline}
\textrm{E}_{\textrm{f},\textrm{V}_\textrm{O}^\textrm{q}}(\epsilon, \mu_\mathrm{O})=\textrm{E}_{\textrm{tot},\textrm{V}_\textrm{O}^\textrm{q}}(\epsilon)-E_{\textrm{tot,stoic}}(\epsilon)+\mu_\mathrm{O} \\
+ \textrm{q} \, [\textrm{E}_\textrm{V}(\epsilon) + \textrm{E}_\textrm{F}(\epsilon)] + \textrm{E}_{\textrm{corr}}(\epsilon) \,,
\label{eq:formenerg}
\end{multline}
where $\epsilon$ is the applied biaxial strain, $\mu_\mathrm{O} = \frac{1}{2}\mu(\textrm{O}_2)+ \Delta \mu(\textrm{O})$ is the oxygen chemical potential with $\mu(\textrm{O}_2)$ being the total energy of an O$_2$ molecule, $\textrm{E}_{\textrm{tot},\textrm{V}_\textrm{O}^\textrm{q}}$ and $\textrm{E}_{\textrm{tot,stoic}}$ are the total energies of the defective and stoichiometric supercells, respectively, $\textrm{E}_\textrm{F}$ is the Fermi energy relative to the valence band maximum ($\textrm{E}_\textrm{V}$) of the defect-free system, which can assume values within the band gap $\textrm{E}_\textrm{g}$ ($0 \leq \textrm{E}_\textrm{F} \leq \textrm{E}_\textrm{g}$) of the stoichiometric structure. We will present results in the oxygen-rich limit, \textit{i.e.} with $\Delta\mu(O)=0$. For charged defects, post-processing of the DFT+$U$ results was performed in order to compute the corrective term $E_{\textrm{corr}}$, necessary to align the electrostatic potential of the defective cell with the one of the neutral defect-free system. This was done by calculating the difference in potential energy between the neutral stoichiometric cell and the charged defective cell via averaging the electrostatic potential in spheres around atomic sites located far from the defect \cite{lany2008}. No further finite-size corrections were applied since the defect concentration we simulate are realistic for this material. 

\section{\label{sec:results}Results and Discussion}

\subsection{Stoichiometric SMO}
\label{sec:results_bulk}

\subsubsection{Empirical versus first-principles $U$ for bulk SMO}
\label{sec:SMObulk_U}

Stoichiometric bulk SMO is often studied in the literature using an empirical ($U_\mathrm{E}$) of about  3.0 eV~\cite{jung1997determination, Hong2012, marthinsen2016coupling}. This value was determined for the chemically similar manganite CaMnO$_3$~\cite{jung1997determination,Hong2012} by reproducing the experimental density of states and the ground-state magnetic order. This approach may present some shortcomings. First of all, the use of the same $U_\mathrm{E}$ in different computational frameworks (different localized Hubbard manifolds, different pseudopotentials, etc.) will not yield the same results for the properties of the same material. Secondly, though CaMnO$_3$ and SrMnO$_3$ are both manganites, they are not structurally identical, and some variations in $U$ for these two materials may be expected~\cite{Hong2012}. Finally, $U_\mathrm{E}$ fitted to one property of the material may not necessarily yield a good description of other properties.

To highlight that $U$ is a Hubbard-manifold and material-dependent property, we have computed the self-consistent Hubbard $U$ for stoichiometric bulk SMO ($U_\mathrm{SC}$) using the DFPT approach of Ref.~\cite{Timrov2018} (see also Sec.~\ref{sec:compdetails}). Results with $U$ values computed from first principles are compared with results obtained by applying a $U$ of 3 eV taken from literature \cite{jung1997determination, Hong2012, marthinsen2016coupling}. We note here that while this value was not determined by fitting to experiment within our computational setup, we still refer it as ``empirical'' ($U_\mathrm{E}$). We have found that for the AFM phase $U_\mathrm{SC}$ is 4.26~eV, while for the FM phase it is 4.36 eV. The computed $U_\mathrm{SC}$ is 1.26 eV larger than $U_\mathrm{E}$ for the AFM phase -- a difference which leads to  changes in the predicted structural, electronic, and magnetic properties of SMO, as we will show in the following.

\begin{table}
\caption{Comparison of the calculated and experimental pseudocubic lattice parameters (\textit{a}, \textit{b}, \textit{c}), volume per formula unit (V), and band gap ($\textrm{E}_\textrm{g}$) of SMO in the most stable AFM phase.}
\begin{tabular*}{\columnwidth}{@{\extracolsep{\fill}}lccccc}
\hline
\hline
Method		 		& \textit{a} (\angstrom) & \textit{b} (\angstrom) & \textit{c} (\angstrom) & V (\angstrom$^3$)	& $\textrm{E}_\textrm{g}$ (eV) \\
\hline
GGA	 		& 3.773 		& 3.771 	 & 3.774	& 53.71 		& 0.44\\
GGA+$U_\mathrm{E}$ 		& 3.796 		& 3.781		 & 3.796	& 54.50 		& 0.57\\
GGA+$U_\mathrm{SC}$ & 3.813 		& 3.790		 & 3.813	& 55.10 		& 0.45\\
Exp. \cite{chmaissem2001relationship}	& 3.805		& 3.805		 & 3.805	& 55.09			& - \\
\hline
\hline
\end{tabular*}
\label{tbl:bulkproperties}
\end {table}

Table~\ref{tbl:bulkproperties} shows the pseudocubic lattice parameters, volume and band gap of AFM SMO, computed using GGA, GGA+$U_\mathrm{E}$ with the empirical Hubbard parameter, and GGA+$U_\mathrm{SC}$ with the self-consistent Hubbard parameter. The experimental lattice parameters of cubic SMO at room temperature \cite{chmaissem2001relationship} are also reported for comparison. We observe that all predicted material parameters are in good agreement with the measured ones, but the error on the lattice parameters is slightly lower when $U_\mathrm{SC}$ is used (mean absolute error on the lattice parameters of 0.8\% for GGA, 0.4\% for GGA+$U_\mathrm{E}$, and 0.3\% for GGA+$U_\mathrm{SC}$). As expected for a nearly cubic system, this is reflected even more clearly in the volume, which is overestimated by only 0.02\% when the structurally consistent $U_\mathrm{SC}$ is applied (compared to -2.50 \% and -1.07 \% at the GGA and GGA+$U_\mathrm{E}$ levels, respectively), in agreement with previous studies using self-consistent $U$ for transition-metal oxides \cite{Hsu2009}. Since experimental lattice parameters were determined at room temperature, we evaluated the lattice thermal expansion within the quasi harmonic approximation~\cite{Togo2010,TOGO20151} and the frozen phonon~\cite{Kresse_1995,TOGO20151} approach. Going from 0 K to room temperature, very small changes in the lattice parameters of about 0.006 and 0.009 \angstrom\ were obtained at the GGA and GGA+$U_\mathrm{SC}$ levels of theory. These changes are smaller (by a factor 2-6) than the difference in lattice parameters reported in Table~\ref{tbl:bulkproperties}  and thus justify the comparison of the 0 K data with the room temperature experimental structure of Ref.~\cite{chmaissem2001relationship}.

The computed $U$ parameters show a small dependence on the magnetic order (about 0.10 eV difference between the AFM and FM phases), which can be explained considering structural and screening effects; where Mn-O distances can be seen as a measure of local screening, since the polarization of neighboring O atoms can effectively screen the Coulomb interaction on the Mn-\textit{3d} orbitals \cite{Lu2014a}. Larger Mn-O distances are obtained at all levels of theory for the FM phase (see Fig. \ref{fig:U_vs_structure} in the supporting information (SI)), which is consistent with the slightly larger $U_\mathrm{SC}$ obtained in this case. A similar picture is obtained from the analysis of the octahedral tilts and rotations which, as the Mn-O bond length, increase linearly with increasing $U$ and which are larger for the FM phase. The reader is referred to Section \ref{sec:bulk_prop_structure} of the SI for further details.

\begin{figure}
 \centering
 \includegraphics[width=0.9\columnwidth]{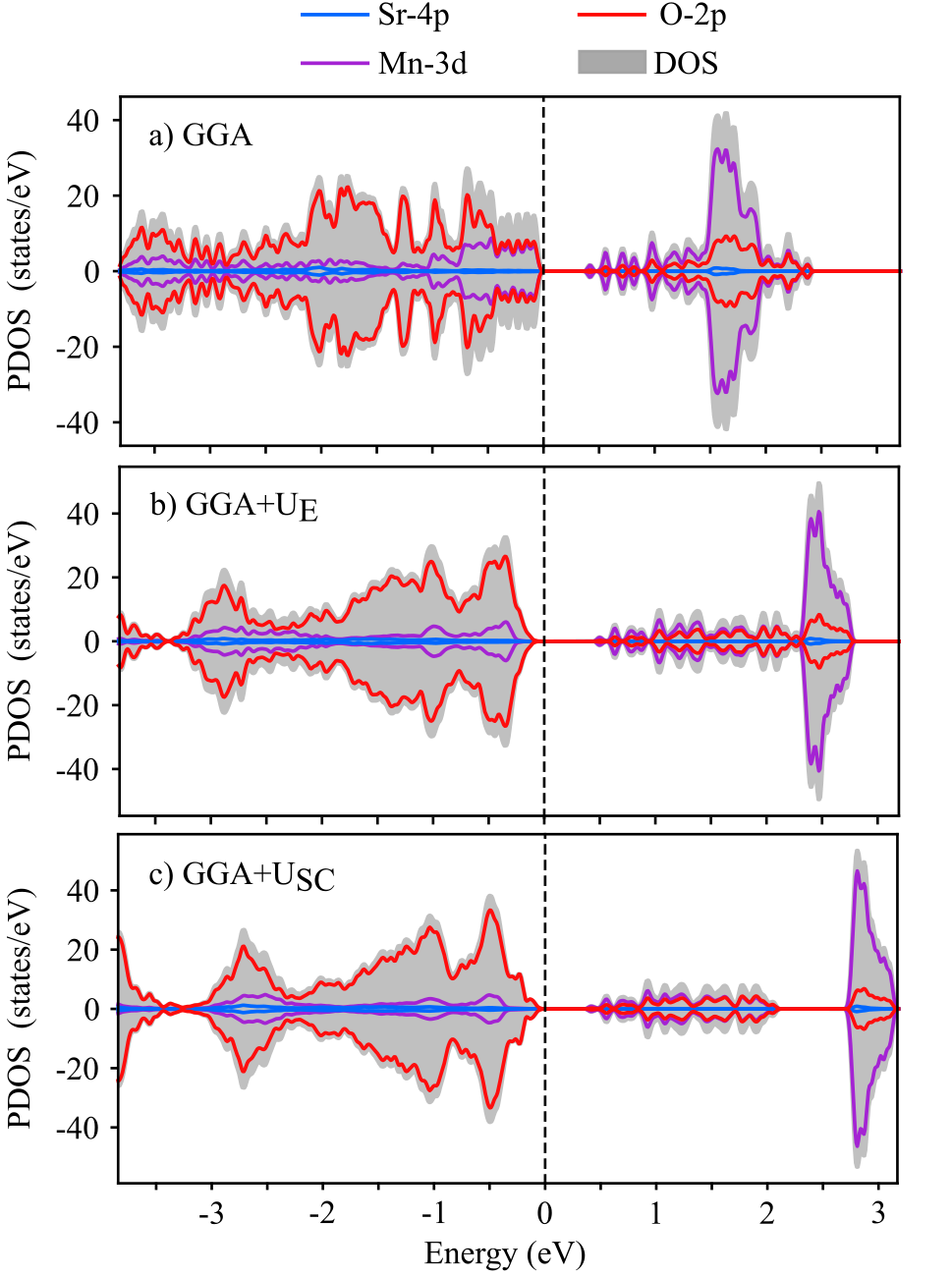}
 \caption{Total and projected density of states (DOS and PDOS, respectively) for bulk SMO in the AFM phase computed using a) GGA, b) GGA+$U_\mathrm{E}$, and c) GGA+$U_\mathrm{SC}$. The zero of the energy scale was aligned with the top of valence band in all cases.}
\label{fig:PDOS_SMO_bulk}
\end{figure}

The differences in computed $U$ values between the two SMO phases can also be explained considering their electronic properties. Table~\ref{tbl:bulkproperties} reports the computed band gap of the ground-state AFM phase of SMO, while the less stable FM phase is found to be metallic at all levels of theory. A small gap of about 0.5 eV is obtained in all cases for the AFM phase. $U_\mathrm{E}$ fitted to the electronic properties of manganites results in the largest band gap. Unfortunately, since the cubic SMO phase is only stable in thin film geometries, the experimental band gap is not available for comparison. The small variation of the band gap as a function of Hubbard $U$ on the Mn-\textit{3d} states was already reported for $U$ varying from 2 to 9 eV~\cite{Ma2010SrMnO3}. Nevertheless, in line with Ref.~\cite{Ma2010SrMnO3}, we observe that the applied $U$ strongly influences the density of states and the contribution of different states to the valence and conduction bands (VB and CB, respectively), as can be seen in Fig.~\ref{fig:PDOS_SMO_bulk}. In particular, at the top of VB we observe a reduction of the Mn-3$d$ character with increasing $U$, while, as expected, empty Mn-\textit{3d} states in the CB are pushed toward higher energies. The shift of CB Mn-\textit{3d} states is however not rigid, the main peak with $t_{2g}$ character shifting more than the $e_g$ states just above the VB maximum. While the combination of these effects results in the observed small variation of the band gap as a function of $U$, the $e_g$ states relevant for defect formation still change significantly for GGA, GGA+$U_\mathrm{E}$ and GGA+$U_\mathrm{SC}$. Unfortunately, in the absence of experimental data, it is not obvious to conclude which $U$ (first-principles or empirical) provides a more accurate picture.

\subsubsection{Strain and magnetic ordering}

Since biaxial strain is known to induce magnetic phase transitions in SMO \cite{lee2010epitaxial, marthinsen2016coupling}, we calculate the total energies of the AFM and FM phases as a function of the applied strain to study the strain-induced transition when using empirical or self-consistent $U$ parameters. When using $U_\mathrm{E}$, the same Hubbard correction is applied to both magnetic orders, while for the self-consistent case we use the respective $U$ values determined above for the AFM and FM phase. Here, we also want to stress that the application of 4\% tensile strain leads to a change in $U_\mathrm{SC}$ of only 0.01~eV compared to zero strain and for simplicity we thus perform epitaxial strain calculations using the same $U$ values as those computed for the unstrained geometries. 

\begin{figure}
 \begin{center}
 \includegraphics[width=0.9\columnwidth]{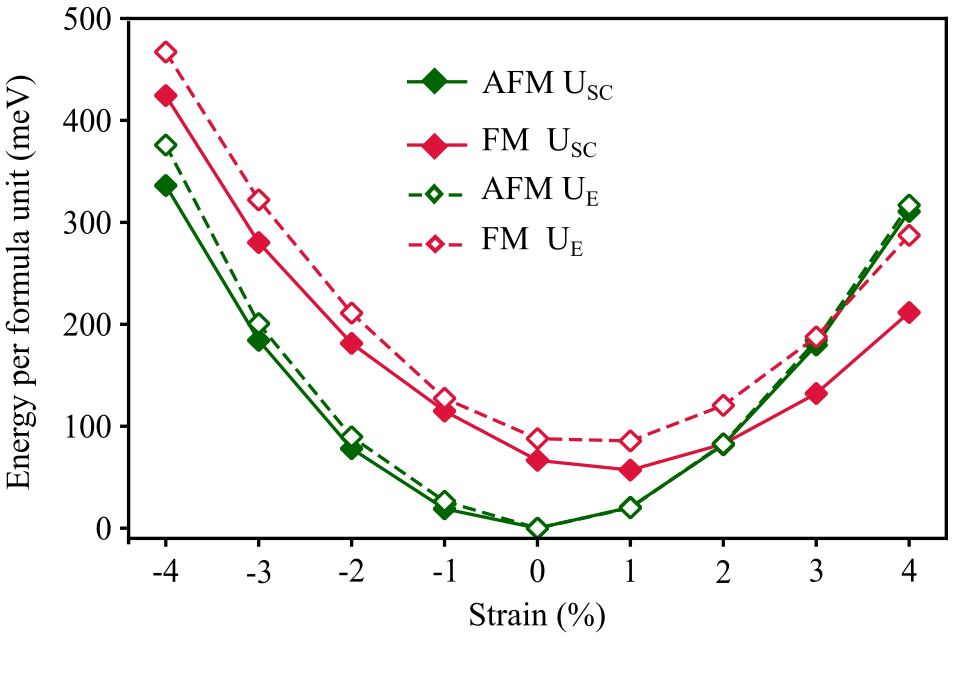}
 \caption{Energy per formula unit as a function of the biaxial strain for AFM and FM magnetic orders computed using GGA+$U_\mathrm{E}$ and GGA+$U_\mathrm{SC}$. Energies are computed relative to AFM 0\% strain. For all strains, the optimized structures belong to space group \textit{Pnma}, except for 4\% compressive strain where the space group is \textit{P4/mbm}.}
\label{fig:strain_bulk}
\end{center}
\end{figure}

In agreement with previous theoretical calculations \cite{lee2010epitaxial, marthinsen2016coupling}, our results (see Fig.~\ref{fig:strain_bulk}) show that with the empirical $U$, the transition from the AFM to the FM ground state is observed at a critical strain of about 3\%. When the self-consistent $U_\mathrm{SC}$ is used instead, the FM order is lower in energy compared to the AFM already for strains of 2\%. The strain-dependent total energy of the AFM phase is similarly described when using both $U_\mathrm{E}$ and $U_\mathrm{SC}$, with small deviations for compressive strain. The FM order instead is consistently predicted to be less stable at the GGA+$U_\mathrm{E}$ level with respect to $U_\mathrm{SC}$ results. This can already be observed for the unstrained structure and becomes even more evident under tensile strain, resulting in the lowering of the critical strain when $U_\mathrm{SC}$ is used.
The larger stabilization of the FM phase is a direct consequence of the larger $U_\mathrm{SC}$ value for this magnetic order. The increased stabilization under tensile strain stems from the fact that as can be seen from Table~\ref{tbl:bulkproperties} and the corresponding $U_\mathrm{E}$ and $U_\mathrm{SC}$ values, the equilibrium volume increases with increasing $U$. This interplay between $U$ and the volume implies that tensile-strained structures with expanded volume are stabilized more when the larger $U_\mathrm{SC}$ for the FM phase is applied. While the $U_\mathrm{E}$ results are in agreement with previous DFT+$U_\mathrm{E}$ calculations and some experiments~\cite{lee2010epitaxial, becher2015strain, maurel2015nature}, comparing the critical strain computed using $U_\mathrm{E}$ or $U_\mathrm{SC}$ with experiments is not straightforward because of the varying concentration of oxygen vacancies in the experimental samples, which, by means of double exchange, could provide a further mechanism for the stabilization of the FM phase as already suggested by experiments~\cite{wang2016oxygen,bai2017structural} and as we will discuss in the following.

\subsection{Defective SMO}
\label{sec:results_defects}

\begin{figure}
\begin{center}
\includegraphics[width=0.9\columnwidth]{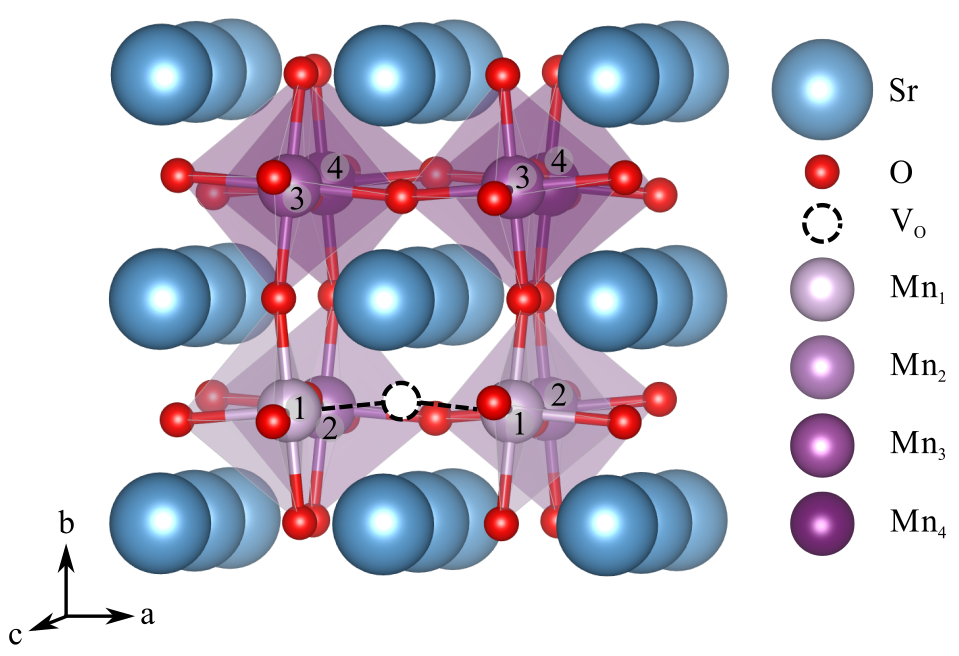}
\caption{Defective SMO supercell with one in-plane oxygen vacancy. Mn atoms are classified in terms of increasing distance from the defect: the larger the index of the Mn atom, the further it is from the defect. The color of the Mn atoms becoming darker with increasing distance, also reflecting their distance from the vacancy.}
\label{fig:SMO_structure_defect}
\end{center}
\end{figure}

Here, we investigate the formation of neutral as well as singly and doubly positively charged oxygen vacancies (V$_\textrm{O}^{\bullet\bullet}$, V$_\textrm{O}^{\bullet}$ and V$_\textrm{O}^{\textrm{X}}$ respectively in  Kr{\"o}ger-Vink notation \cite{KROGER1956307}, see also SI Section \ref{sec:defect_types}) in SMO bulk and strained thin-films. While in the stoichiometric bulk structure all Mn atoms are equivalent (see Fig.~\ref{fig:SMO_structure_bulk}) and thus have the same Hubbard $U$, the creation of a defect breaks the symmetry and Hubbard sites at different distances from the defect will have different chemical environments. In particular, upon oxygen-vacancy creation, Mn atoms in nearest-neighbor positions (Mn$_1$ in Fig.~\ref{fig:SMO_structure_defect}) have a different coordination number due to the broken Mn$_1$--O--Mn$_1$ bond. Depending on the charge state of the defect we also expect changes in oxidation state for the Mn$_1$ ions, as outlined in the SI Section \ref{sec:defect_types}. Besides these apparent modifications we also expect a perturbation of the chemical environment of the atoms further from the vacancy, that decays with increasing Mn--V$_\textrm{O}$ distance (see Fig.~\ref{fig:SMO_structure_defect}). This is expected to lead to different Hubbard $U$ parameters of inequivalent Mn sites and that a self-consistent site-dependent Hubbard $U$ approach that accounts for these local chemical effects should be applied to properly describe the properties of defective SMO. In the following, we will show how SC-SD $U$ values influence the predicted defect formation energies, structural and electronic properties, and magnetic phase transitions for oxygen vacancies in SMO.

\subsubsection{Self-consistent site-dependent $U$ parameters}

\begin{figure}[t]
 \centering
 \includegraphics[width=0.9\columnwidth]{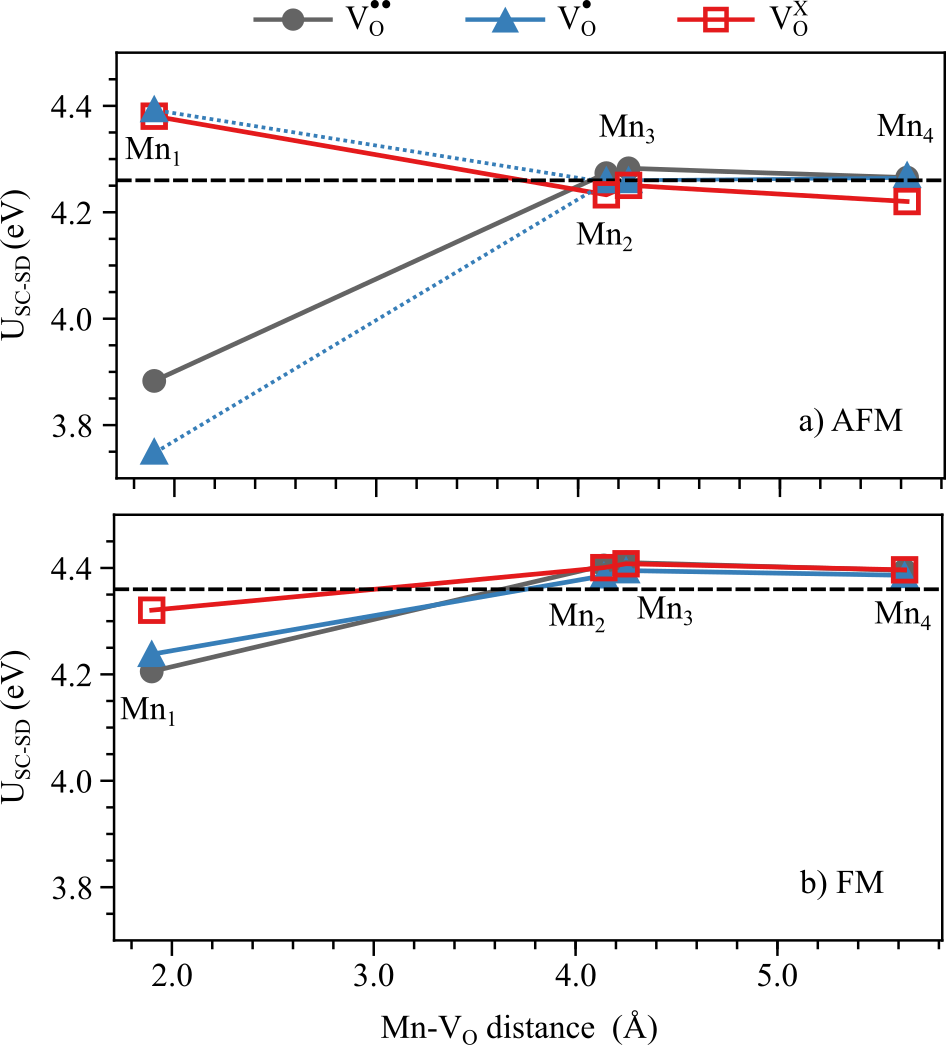}
 \caption{Changes in $U_\mathrm{SC-SD}$ of Mn sites in bulk SMO with a V$_\textrm{O}^{\bullet\bullet}$, V$_\textrm{O}^{\bullet}$, or V$_\textrm{O}^{\textrm{X}}$ defect for the a) AFM and b) FM phase. Dashed horizontal lines indicate the respective $U_\mathrm{SC}$ (\textit{i.e.} no oxygen vacancy). See also Fig. \ref{fig:SMO_structure_defect} for the Mn labels. Solid and dotted lines are guides to the eye. In panel a) there are two $U_\textrm{SC-SD}$ values for Mn$_1$ (see the main text for details).}
\label{fig:usc_defects}
\end{figure}

We compute self-consistent site-dependent Hubbard $U$ parameters ($U_\mathrm{SC-SD}$, see Sec.~\ref{sec:compdetails}), for all inequivalent Mn atoms in SMO with oxygen vacancies in different charge states (see above) and with different magnetic orders (FM and AFM). We find that $U_\mathrm{SC-SD}$ depends on three factors: the charge state of the oxygen vacancy, the magnetic order of the defective SMO, and the distance of Mn sites from the oxygen vacancy, as can be seen by comparing the results in Figs.~\ref{fig:usc_defects}~(a) and (b). Similarly to the stoichiometric material, $U$ values are larger for the FM order also in defective SMO. As expected from the above discussion, we observe deviations from $U_\mathrm{SC}$, computed for the stoichiometric bulk, mainly for Mn$_1$ in nearest-neighbor position to the defect at a distance of about 1.90 \angstrom, while Mn sites at larger distances recover $U_\mathrm{SC}$ (the deviations are as small as 0.04 eV). For this reason we will restrict our discussion to $U_\mathrm{SC-SD}$ of the Mn$_1$ site ($U_\mathrm{Mn_1}$).

For the AFM phase (see Fig.~\ref{fig:usc_defects}a) we observe a small change (+0.11~eV) of $U_\mathrm{Mn_1}$ with respect to $U_\mathrm{SC}$ for a V$_\textrm{O}^{\textrm{X}}$. This is in line with the absence of changes in the Mn$_1$ oxidation state after removal of the O$^{2-}$ anion and we associate the increase in $U_\mathrm{Mn_1}$ to the change in coordination number. A larger deviation (-0.38~eV) is observed for V$_\textrm{O}^{\bullet\bullet}$, reflecting both the change in coordination number and oxidation state when both Mn$_1$ sites are reduced.  V$_\textrm{O}^{\bullet}$ is an intermediate case and we observe different $U_\mathrm{SC-SD}$ on the two Mn$_1$ sites: one $U$ is much smaller (-0.52~eV) compared to $U_\mathrm{SC}$, while the other is slightly larger (+0.13~eV). This latter value is very close to the one observed for the V$_\textrm{O}^{\textrm{X}}$ suggesting that only one of the two Mn$_1$ is reduced, while the second retains its Mn$^{4+}$ oxidation state. For a V$_\textrm{O}^{\bullet}$ the two Mn$_1$ thus become chemically inequivalent despite having the same coordination and distance from the defect. Based on these observation we believe that a loss in coordination leads to a slightly increased $U$, while the reduction is reflected by a more marked decrease of $U$. Quantitatively explaining the change in $U$ with the oxidation state of an ion is in general not straightforward. While $d$ states do not contribute to screening (which is performed by the other, more delocalized, states in the crystal), the change of the number of $d$ electrons in one ion can certainly change its electronic structure (\textit{e.g.}, contract orbitals towards the nucleus) or determine the number of ligands, the distance from neighbor anions or the ionicity of metal-ligand bonds, which are all factors able to influence the value of $U$. The LR-DFPT calculations of $U$ used here ensure that the value of the Hubbard parameter is consistent with all these factors and are capable to guarantee a more sound representation of the behavior of the metal ions, consistently with their oxidation state and crystal environment~\cite{Cococcioni2018}.

\begin{figure}
 \centering
 \includegraphics[width=0.9\columnwidth]{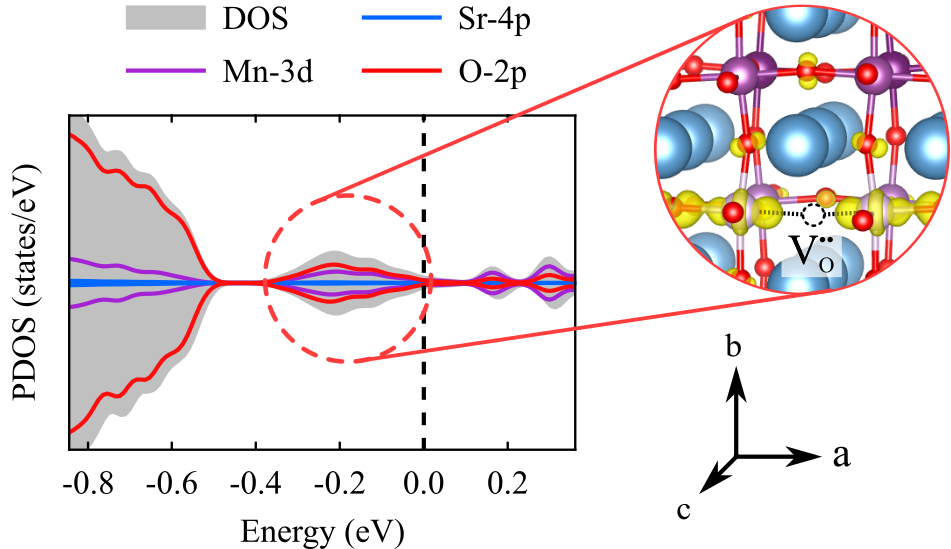}
 \caption{Total and projected density of states (DOS and PDOS, respectively) for a V$_\mathrm{O}^{\bullet \bullet}$ in the AFM phase of SMO computed using $U_\mathrm{SC-SD}$. The vertical dotted line indicates the position of the Fermi level. The isosurface (10$^{-2}$ \textit{e}\angstrom $^{-3}$) in the inset shows the density associated with the circled defect state. The vacancy position is indicated by the dashed circle in the inset.}
\label{fig:PDOS_SMO_nVO_SCSD}
\end{figure}

The above interpretation of the electronic structure and resulting changes in $U_\mathrm{SC-SD}$ is supported by an analysis of the density of states and oxidation states. As shown in Fig. \ref{fig:PDOS_SMO_nVO_SCSD} at the $U_\mathrm{SC-SD}$ level the defect state is well separated from the VB and CB (see SI Section \ref{sec:defect_elprop} for a comparison of the defect state at different $U$ values). The inset shows the electronic density associated with the defect state, which is mainly localized on the two Mn$_1$ with minor contributions of the neighboring oxygen atoms. This highlights the reduction of the two Mn$_1$ from Mn$^{4+}$ to Mn$^{3+}$ leading to large chemical perturbations reflected in $U_\mathrm{SC-SD}$. A similar interpretation is provided by the oxidation states computed from occupation numbers of the Mn-$3d$ states \cite{sit2011simple,walsh2018oxidation}. From the oxidation states reported in Table \ref{tbl:OS} (see also Table \ref{tbl:SIos} in the SI for more details) we clearly see that both an empirical $U_\mathrm{E}$ and $U_\mathrm{SC-SD}$ result in the same reduction of the two Mn$_1$ for V$_\textrm{O}^{\bullet\bullet}$, which is however absent at the pure GGA level (see SI Section \ref{sec:defect_elprop} for a discussion of the DOS at various $U$ values). For the V$_\textrm{O}^{\bullet}$ we observe reduction of one of the two Mn$_1$ independent from $U$, while both Mn$_1$ retain their 4+ oxidation state for the V$_\textrm{O}^{\mathrm{X}}$, in agreement with the above discussion.

For the FM phase (see Fig.~\ref{fig:usc_defects}b), $U$ for the Mn$_1$-$3d$ states ($U_\mathrm{Mn_1}$) is found to deviate less from $U_\mathrm{SC}$ compared to the AFM phase. As expected from the above discussion for the AFM phase, changes in $U_\mathrm{Mn_1}$ decrease in order of V$_\textrm{O}^{\bullet\bullet}$ (-0.15~eV), V$_\textrm{O}^{\bullet}$ (-0.12~eV), and V$_\textrm{O}^{\textrm{X}}$ (-0.04~eV). Interestingly, the two Mn$_1$ atoms next to V$_\textrm{O}^{\bullet}$ now show the same behavior, even if the structural symmetry is artificially broken in order to differentiate the two sites. This observation can be related to the metallic nature of FM SMO, in which we expect the defect states (and the changes in the chemical environment due to their formation) to be more delocalized over the whole structure, compared to the semiconducting AFM order, where localized defect states in the band gap are expected to lead to a more confined impact on the local chemical environment as we will discuss in more detail below.

\begin{table}
\caption{Oxidation states for Mn atoms in stoichiometric and defective AFM bulk SMO with different $U$ parameters. See Fig. \ref{fig:SMO_structure_defect} for the Mn labels. For V$^{\bullet}_\mathrm{O}$, the two Mn$_1$ atoms show different behaviors and are hence reported separately (Mn$_1$(a) and Mn$_1$(b)).}
\begin{tabular*}{\columnwidth}{@{\extracolsep{\fill}}llccc}
\hline
\hline
Vacancy type		 	& Site & GGA & GGA+$U_\mathrm{E}$ & GGA+$U_\mathrm{SC-SD}$\\
\hline
\multirow{2}{*}{V$_\textrm{O}^{\bullet \bullet}$} & Mn$_1$ & 4+ & 3+ & 3+\\
												  & Mn$_{2-4}$ 	& 4+ & 4+ & 4+\\
\hline
\multirow{3}{*}{V$_\textrm{O}^{\bullet}$} 		  & Mn$_1$(a) & 3+ & 3+ & 3+\\
												  & Mn$_1$(b) & 4+ & 4+ & 4+\\
												  & Mn$_{2-4}$ 	& 4+ & 4+ & 4+\\
                                                  \hline
\multirow{2}{*}{V$_\textrm{O}^{\textrm{X}}$} 	  & Mn$_1$ & 4+ & 4+ & 4+\\
												  & Mn$_{2-4}$ 	& 4+ & 4+ & 4+\\
\hline
Stoichiometric								  & Mn			& 4+ & 4+ & 4+\\
\hline
\hline
\end{tabular*}
\label{tbl:OS}
\end {table}

\subsubsection{Defect formation energies and magnetic phase transitions}

\begin{figure}[t]
 \centering
 \includegraphics[width=0.9\columnwidth]{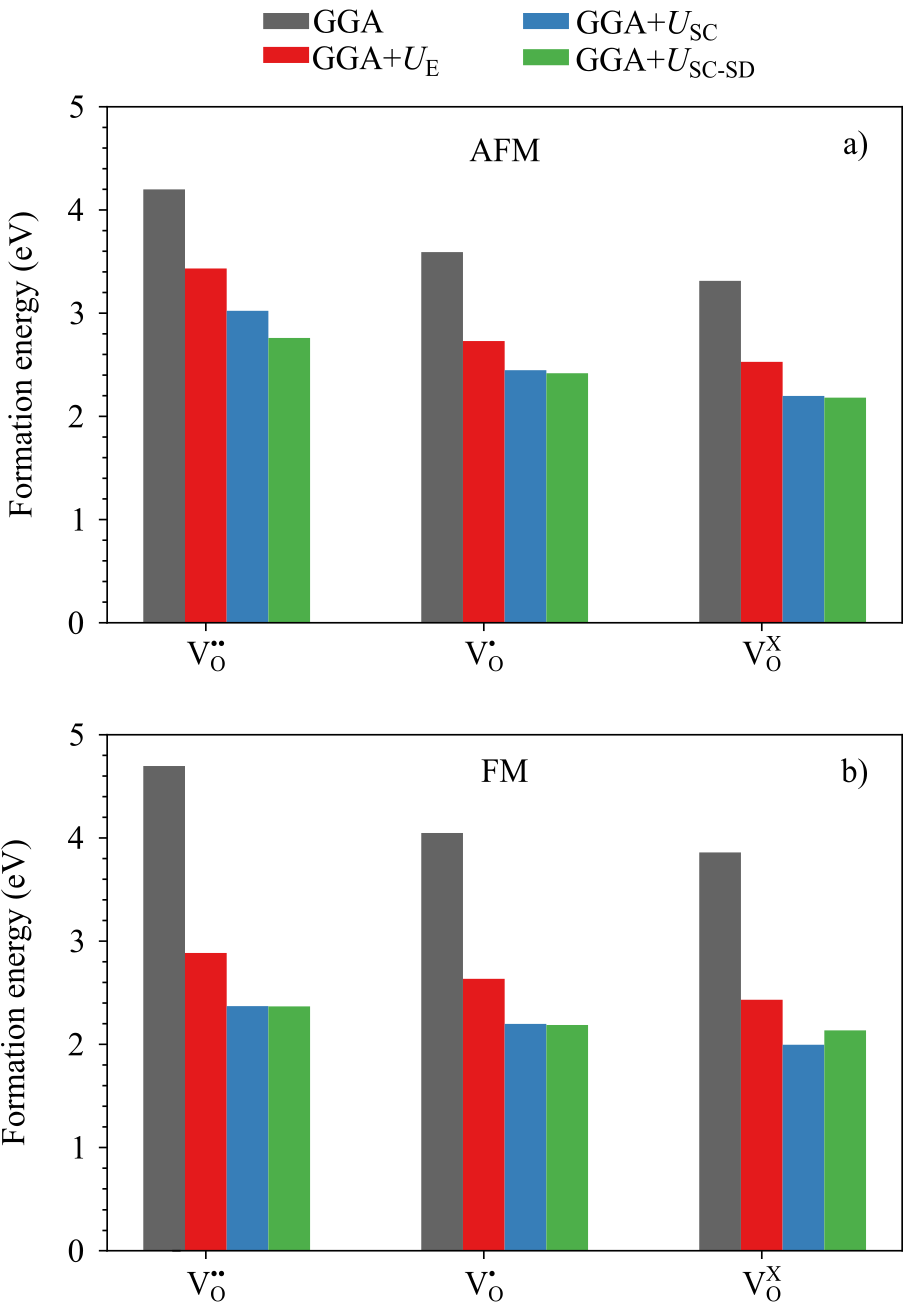}
 \caption{Formation energies computed for V$_\textrm{O}^{\bullet \bullet}$, V$_\textrm{O}^{\bullet}$ and V$_\textrm{O}^\textrm{X}$ in SMO using different methods for $\Delta \mu(\mathrm{O}) = 0$ and E$_\mathrm{F} = 0$ in a) the AFM and b) the FM phase.}
\label{fig:form_energ}
\end{figure}

Taking into account changes in $U$ for Mn sites around the defect does not significantly alter the description of the structural and electronic properties compared to $U_\mathrm{SC}$, as can been seen in more detail in Sections \ref{sec:defect_structure} and \ref{sec:defect_elprop} of the SI. However, as we will show in the following, the $U_\mathrm{SC-SD}$ approach leads to important quantitative changes in the formation energies and magnetic properties of defective SMO. Including local chemical changes on the transition-metal atoms around the defect is thus important for all properties related to defect energetics, as for example defect-induced magnetic phase transitions.

Fig.~\ref{fig:form_energ} shows the formation energies for V$_\textrm{O}^{\bullet \bullet}$, V$_\textrm{O}^{\bullet}$ and V$_\textrm{O}^\textrm{X}$ defects in bulk SMO, comparing results computed with different $U$ values for both the AFM and FM magnetic orders. Independently of the defect charge or the magnetic phase, DFT+$U$ lowers the formation energy with respect to GGA, reflecting the known issues related to the incorrect electronic-structure description of semi-local functionals for defective transition metal oxides. Comparing results for $U_\mathrm{E}$ and $U_\mathrm{SC}$, we can see that the formation energies of oxygen vacancies strongly depend on the applied $U$. This suggests that using an empirical $U$ that correctly reproduces the experimental electronic structure of the bulk (an incorrect requirement even for exact DFT) does not necessarily lead to the best description of the defective system. The effect of the site-dependent $U_\mathrm{SC-SD}$ approach on the formation energy quite strongly depends on the defect charge and on the magnetic order. In line with the magnitude of changes of $U_\mathrm{SC-SD}$, differences in formation energies between GGA+$U_\mathrm{SC-SD}$ and GGA+$U_\mathrm{SC}$ are larger for the AFM than the FM phase, which stems from the localized defect states in the former. The difference decreases with increasing charge of the defect, reaching negligible levels (-0.02 eV) for the V$_\textrm{O}^{\mathrm{X}}$ in the AFM phase, while even being positive in the FM phase.

\begin{figure}
 \centering
 \includegraphics[width=0.9\columnwidth]{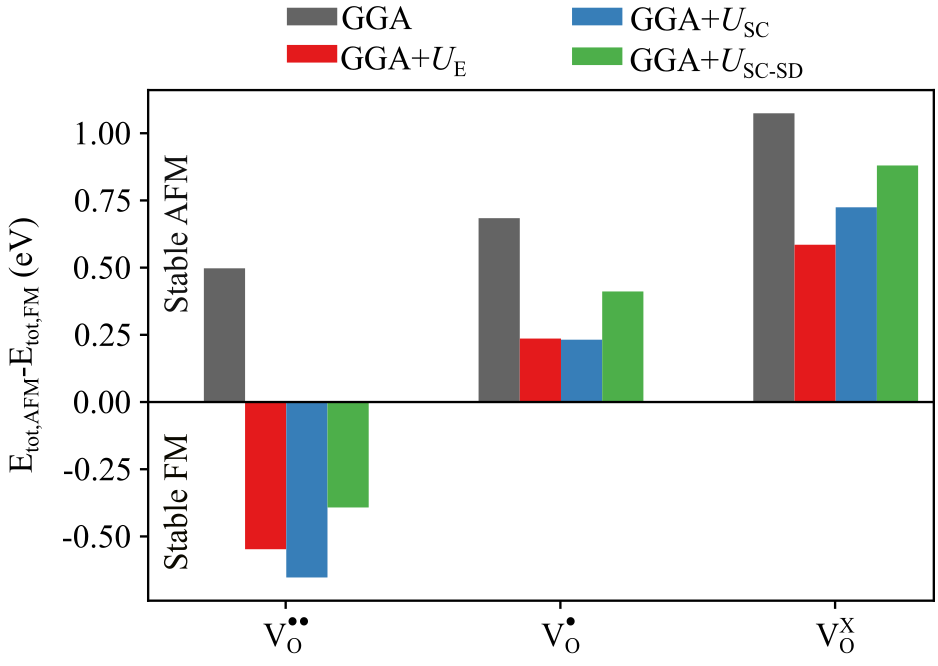}
 \caption{Total energy differences between defective (V$_\textrm{O}^{\bullet \bullet}$, V$_\textrm{O}^{\bullet}$ and V$_\textrm{O}^\textrm{X}$) cells with AFM and FM order, computed with different methods. AFM is more stable for positive and FM for negative differences.}
\label{fig:fm_vs_afm_defect}
\end{figure}

Using DFT+$U_\mathrm{E}$ calculations, a 4.2\% oxygen vacancy concentration in SMO was shown to lead to a magnetic phase transition from AFM to FM~\cite{marthinsen2016coupling}. To assess the effect of $U_\mathrm{SC-SD}$ on this phase transition, we show in Fig.~\ref{fig:fm_vs_afm_defect} the total-energy difference between defective cells with FM and AFM magnetic orders computed using different methods. For V$_\textrm{O}^{\bullet \bullet}$, GGA predicts the AFM order to be more stable, but increasing $U$ from 0 to 3 ($U_\mathrm{E}$) or about 4 eV ($U_\mathrm{SC}$), the FM order is increasingly favored. This energetic preference can be rationalized by the small energetic cost associated with accommodating the two excess electrons on delocalized Mn states in the metallic FM phase. When using $U_\mathrm{SC-SD}$, Mn$_1$ atoms assume lower $U$ values compared to $U_\mathrm{SC}$ or to Mn sites farther away from the defect (see Fig.~\ref{fig:usc_defects}) in both the AFM and FM orders. We saw above that increasing $U$ favors the FM order, which conversely implies that the decreased $U$ on Mn$_1$ will locally destabilize the FM order as can be seen by the 0.26 eV smaller total energy difference for $U_\mathrm{SC-SD}$ compared to $U_\mathrm{SC}$. Even though the AFM order is favoured for the charged defects, independently of the method, we see a similar effect when comparing $U_\mathrm{SC-SD}$ to $U_\mathrm{SC}$. The former leads to a higher stability of the AFM phase by 0.18 eV and 0.15 eV respectively for V$_\textrm{O}^{\bullet}$ and V$_\textrm{O}^\textrm{X}$. While all DFT+$U$ methods thus yield qualitatively the same magnetic phase stability, the quantitative differences between $U_\mathrm{SC}$ and $U_\mathrm{SC-SD}$ will be important when assessing magnetic transition temperatures.

\subsubsection{Interplay between defects and applied strain}

\begin{figure}
 \centering
 \includegraphics[width=0.9\columnwidth]{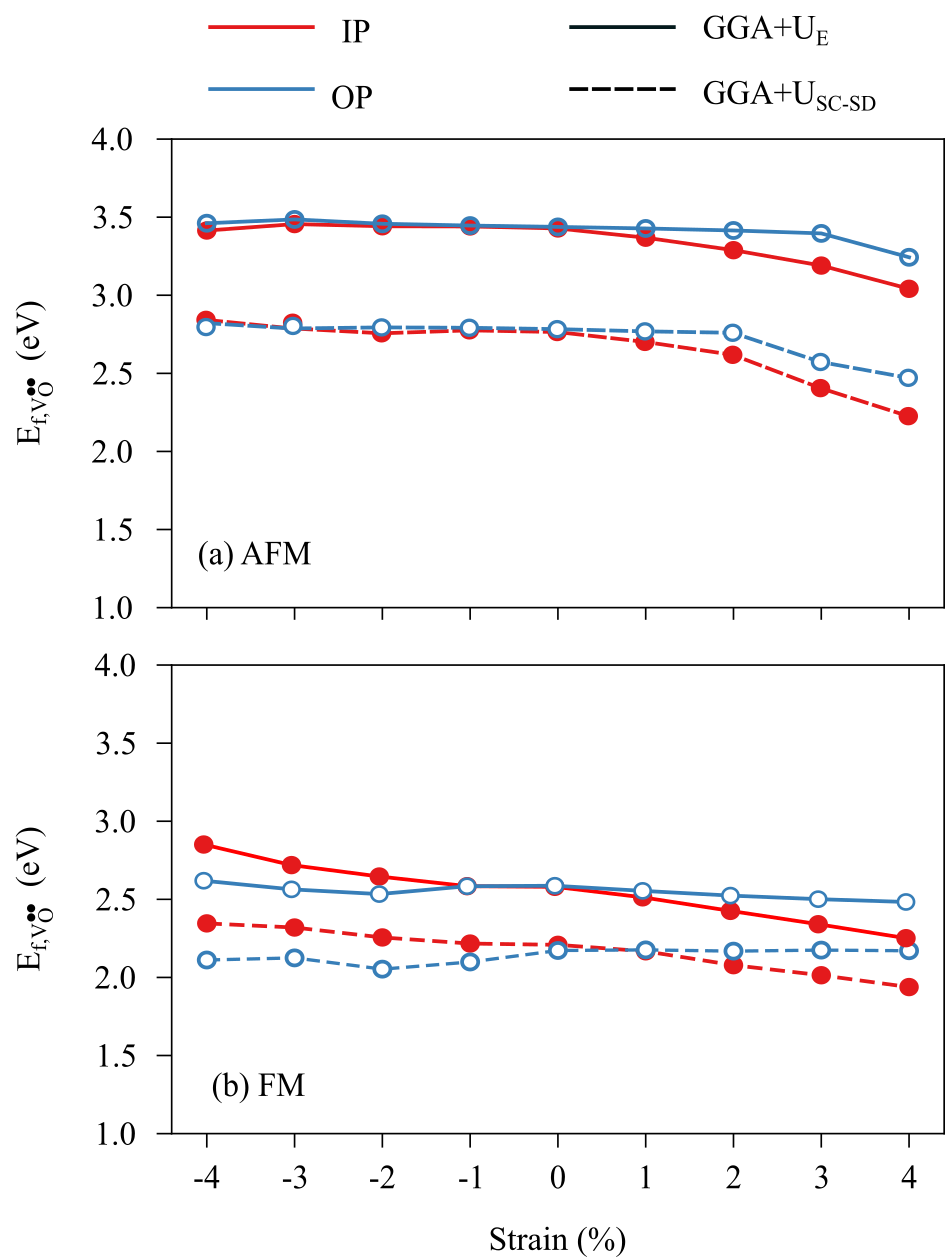}
 \caption{Strain dependence of the formation energy of an in-plane (IP, broken Mn-O-Mn bond in the biaxial strain plane) and out-of-plane (OP, broken Mn-O-Mn bond perpendicular to the biaxial strain plane) neutral oxygen vacancy computed using  $U_\mathrm{E}$ and $U_\textrm{SC-SD}$ in a) the AFM and b) the FM phase of SMO. For all defective structures, the space group is \textit{P1}.}
\label{fig:strained_vac}
\end{figure}

In thin film geometries, the oxygen-vacancy formation energy and consequently the defect concentration depend on volume changes induced by biaxial strain\cite{aschauer2013strain, aschauer2016interplay, marthinsen2016coupling}. Moreover, biaxial strain breaks the symmetry \cite{Rondinelli:2011jk} and thus allows strain-controlled ordering of defects on inequivalent sites \cite{marthinsen2016coupling, aschauer2013strain, aschauer2016interplay}. In this section, we discuss how the use of SC-SD DFT+$U$ affects the interplay between strain, oxygen vacancies and magnetism in SMO. For the sake of simplicity, we will concentrate in the comparison of results obtained with $U_\mathrm{E}$ as in previous works\cite{marthinsen2016coupling} and with $U_\mathrm{SC-SD}$. Fig. \ref{fig:strained_vac} shows strain-dependent formation energies of neutral oxygen vacancies both on the IP site (broken Mn-O-Mn bond in the biaxial strain plane) and the OP site (broken Mn-O-Mn bond along the film normal) of SMO computed using $U_\mathrm{E}$ or $U_\mathrm{SC-SD}$ for both the AFM and the FM magnetic orders. In analogy to the above observations for bulk SMO, formation energies computed at the GGA+$U_\mathrm{SC-SD}$ level are lower compared to GGA+$U_\mathrm{E}$ over the whole strain range. Also in analogy to the bulk, we can see that $U_\mathrm{SC-SD}$ results in smaller differences between formation energies in the AFM and FM phase compared to $U_\mathrm{E}$ as the average difference between formation energies in the two phases is reduced from about 1 eV with $U_\mathrm{E}$ to about 0.5 eV with $U_\mathrm{SC-SD}$. With $U_\mathrm{SC-SD}$ we moreover observe that the formation energy in the AFM phase approaches that in the FM phase, suggesting that tensile strain values larger than the ones computed here should lead to a stable AFM magnetic order in defective SMO.

4\% tensile strain decreases the formation energy of an IP vacancy in the AFM phase by about 0.5 eV for $U_\mathrm{SC-SD}$ and 0.4 eV for $U_\mathrm{E}$, corresponding respectively to 18\% and 12\% of the unstrained bulk value. Strain thus strongly alters the equilibrium oxygen content, but the two approaches predict different changes in the defect concentration. Interestingly, if the $U_\mathrm{SC}$ is used instead of $U_\mathrm{SC-SD}$, we find the same reduction of 0.4 eV as for $U_\mathrm{E}$. The additional stabilization of the neutral oxygen vacancy with $U_\mathrm{SC-SD}$ thus stems from the marked changes of $U$ on Mn$_1$ sites that is a result of properly taking into account the local chemical environment of the Hubbard sites upon defect formation.  Interestingly, the cooperative effect of strain and defects can be clearly seen by the fact that both $U_\mathrm{E}$ and $U_\mathrm{SC-SD}$ predict a change in the slope of the formation energy for the defective AFM systems exactly at the critical strain for the AFM to FM transition (2\% for $U_\mathrm{SC-SD}$ and 3\% for $U_\mathrm{E}$): the increased stabilization of the defect for larger strains reflects the reduced thermodynamic stability of the AFM phase.

For the FM phase, changes in $U$ are smaller, which is reflected in the similarity of the strain dependence of the neutral oxygen vacancy formation energy when using $U_\mathrm{E}$ and $U_\mathrm{SC-SD}$. We also note that for the FM phase, we observe the increase in IP formation energy expected from volume arguments, which is not observable in the AFM phase due to crystal field effects \cite{aschauer2013strain}. In the FM phase, the metallicity leads to a reduced sensitivity of the total energy on these crystal-field changes and volume effects dominate.

\section{\label{sec:concls}Conclusions}

In the present work we have applied the DFT+$U$ approach using either empirical Hubbard $U$ parameters ($U_\mathrm{E}$) or with $U$ parameters computed self-consistently ($U_\mathrm{SC}$) for bulk SMO, or self-consistently and site-dependently ($U_\mathrm{SC-SD}$) for SMO with oxygen vacancies. The DFT$+U_\mathrm{SC-SD}$ approach allows to properly account for both structural and local chemical effects through the determination of $U$ from first principles for all inequivalent Hubbard sites in the system. The site-dependence of $U$ allows to properly describe the excess charge localization and coordination changes in the structure upon defect creation, while the self-consistent procedure ensures the required internal consistency of the results. 

For the stoichiometric bulk material, where all the Hubbard sites are equivalent, the $U$ values show a small dependence on the magnetic phase of the system, with differences that can be explained in terms of local screening (Mn-O bond length) and other structural effects (i.e. octahedral rotations). $U_\mathrm{SC}$ improves the prediction of lattice parameters compared to GGA or GGA+$U_\mathrm{E}$ and strongly affects states involved in defect formation, both of which are crucial for the prediction of strain and defect induced changes in properties.

In defective systems, the formation of O vacancies induces changes in the chemical environment around the defect, resulting in inequivalent Hubbard atoms. This is directly reflected in their $U_\mathrm{SC-SD}$ values, which are affected by the distance of the site from the defect, its coordination number, oxidation state and the defect charge as well as the magnetic order of the bulk. Site-dependent changes in $U$ are found to be more pronounced for the semi-conducting AFM phase than for the metallic FM phase, which we can relate to the localized defect state in the former that leads to stronger chemical alterations around the defect site. This is also reflected in the decrease of the variation in $U$ for charged defects, where no reduction of sites around the defect occurs. While the global structural and electronic properties are not significantly affected by $U_\mathrm{SC-SD}$ compared to $U_\mathrm{SC}$, the site dependence of $U$ has a strong impact on the computed formation energies and consequently on all the properties related to the defect energetics, such as the energetic preference of the defect-induced AFM$\rightarrow$FM phase transition or strain-induced vacancy ordering. 

The current study demonstrates that DFT+$U_\mathrm{SC-SD}$ is a promising approach to study the energetics of defects in semiconducting or insulating transition-metal oxides, where defects may lead to filled localized states in the band gap. Future work should focus for example on the applicability of the approach to shallow defect states where we expect a more long-range dependence of the site-dependent $U$ parameters due to the larger delocalization of defect states.

\section*{\label{sec:acknow}Acknowledgments}

This research was supported by the NCCR MARVEL, funded by the Swiss National Science Foundation. Computational resources were provided by the University of Bern (on the HPC cluster UBELIX, http://www.id.unibe.ch/hpc), by the Swiss National Supercomputing Center (CSCS) under projects ID mr26 and s836 and SuperMUC at GCS@LRZ, Germany, for which we acknowledge PRACE for awarding us access.

\bibliography{references}


\clearpage
\clearpage 
\setcounter{page}{1}
\renewcommand{\thetable}{S\arabic{table}}  
\setcounter{table}{0}
\renewcommand{\thefigure}{S\arabic{figure}}
\setcounter{figure}{0}
\renewcommand{\thesection}{S\arabic{section}}
\setcounter{section}{0}
\renewcommand{\theequation}{S\arabic{equation}}
\setcounter{equation}{0}
\onecolumngrid

\begin{center}
\textbf{Supplementary information for\\\vspace{0.5 cm}
\large Self-consistent site-dependent DFT+$U$ study of stoichiometric and defective SrMnO$_3$\\\vspace{0.3 cm}}
Chiara Ricca,$^{1, 3}$ Iurii Timrov,$^{2, 3}$ Matteo Cococcioni,$^{2, 3}$ Nicola Marzari,$^{2, 3}$ and Ulrich Aschauer$^{1, 3}$

\small
$^1$\textit{Department of Chemistry and Biochemistry, University of Bern, Freiestrasse 3, CH-3012 Bern, Switzerland}

$^2$\textit{Theory and Simulation of Materials (THEOS), Ecole Polytechnique F\'ed\'erale de Lausanne, CH-1015 Lausanne, Switzerland}

$^3$\textit{National Centre for Computational Design and Discovery of Novel Materials (MARVEL), Switzerland}

(Dated: \today)
\end{center}

\section{\label{sec:bulk_prop} Stoichiometric SMO}

\subsection{\label{sec:bulk_prop_structure} Crystal Structure}
\begin{figure}[h]
 \centering
 \includegraphics{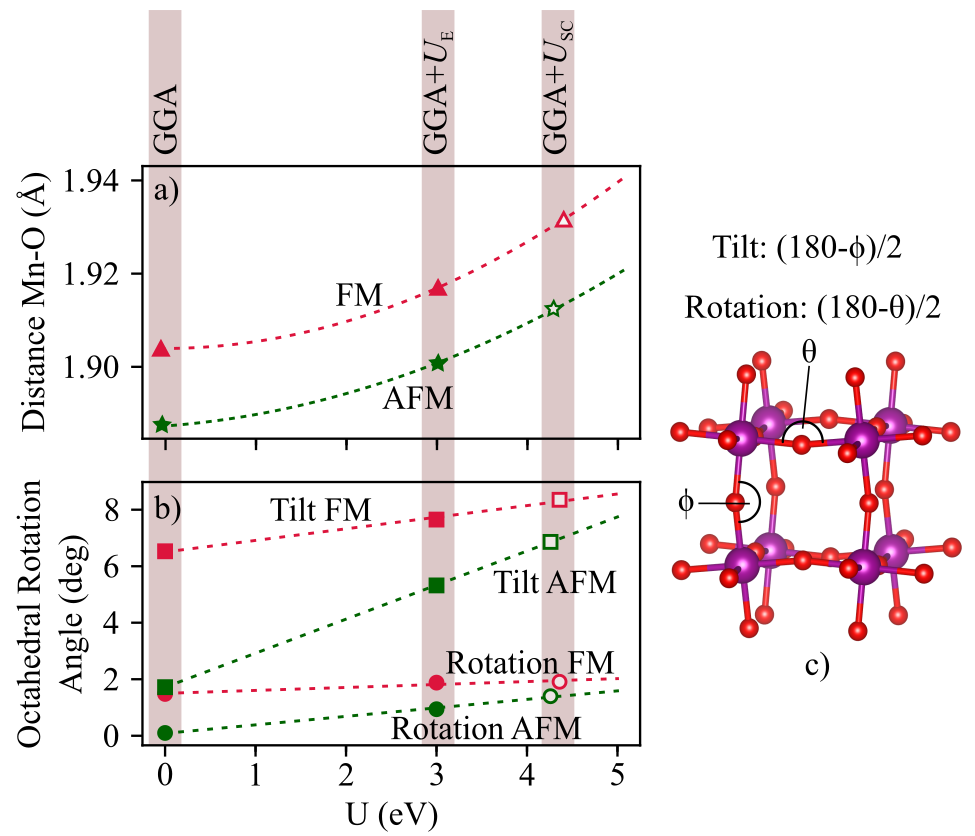}
 \caption{Evolution of a) the average Mn-O bond distance and b) the octahedral rotation (Mn-O-Mn angle in the \textit{ac} plane, see panel c) and tilts (Mn-O-Mn angle along the \textit{b} axis, see panel c) as a function of $U$ values applied to the AFM (in red) or FM (in green) SMO phases. Hollow markers represent data obtained using $U_\mathrm{SC}$  for the two phases. Dotted lines are quadratic and linear fits in a) and b) respectively. c) Definition for the rotation ($\theta$) and tilt ($\phi$) angles. }
\label{fig:U_vs_structure}
\end{figure}

Mn-O distances as well as octahedral tilt, and rotation angles, as defined in Fig. \ref{fig:U_vs_structure}c, can  be used to characterize the effect of Hubbard $U$ on the structure. Mn-O distances can be seen as a measure of the local screening since the polarization of the neighboring O atoms can effectively screen the Coulomb interaction on the Mn-\textit{3d} orbitals \citeSI{Lu2014aSI}. In Fig.~\ref{fig:U_vs_structure}a we can see a quadratic-like behavior of the average Mn-O bond length as a function of $U$, with larger $U$ values resulting in longer bonds, as a result of a weaker screening. This is also consistent with the increase of the volume observed when increasing $U$ (see Table \ref{tbl:bulkproperties} in the main text). Larger Mn-O distances are obtained at all levels of theory for the FM phase (see Fig.~\ref{fig:U_vs_structure}a), which is consistent with the slightly larger $U_\mathrm{SC}$ value obtained in this case and the weaker screening of this metallic phase. A similar picture is obtained from the analysis of the octahedral tilts and rotations, which increase linearly with increasing $U$ and are larger for the FM phase (see Fig.~\ref{fig:U_vs_structure}b). Usually increasing the volume tends to eliminate octahedral rotations\citeSI{Rondinelli:2011jkSI}. The present increase in octahedral rotation angles despite the volume increase for larger $U$ values can be explained by the Bader atomic volumes~\citeSI{bader1991quantumSI}, which also increase as the Hubbard parameter goes from 0 eV to $U_\mathrm{SC}$ (cf. Table \ref{tbl:bulk_volumes}).

\clearpage

\begin{table}[!h]
\caption{Bader atomic volumes (V, in a.u.$^3$) for Sr, Mn, and O in stoichiometric SMO computed at different levels of theory.}
\begin{tabular*}{\columnwidth}{@{\extracolsep{\fill}}llccc}
\hline
\hline
Phase & Method & V$_{Sr}$	& V$_{Mn}$ &V$_{O}$\\ 
\hline
    & GGA 					& 98.12 & 45.64 &	72.89\\
AFM & GGA+$U_\mathrm{E}$ 	& 98.29 & 46.32 &	74.41	\\
    & GGA+$U_\mathrm{SC}$ 	& 98.59 & 47.02 &	75.38 	\\		
\hline
    & GGA 					& 97.88  & 46.89 &	74.86\\
 FM & GGA+$U_\mathrm{E}$ 		& 98.77  & 47.45 &	76.00	\\
    & GGA+$U_\mathrm{SC}$ 	& 100.16 & 48.43 &	77.73 	\\
\hline
\hline
\end{tabular*}
\label{tbl:bulk_volumes}
\end {table}

\begin{figure}[h]
 \centering
 \includegraphics{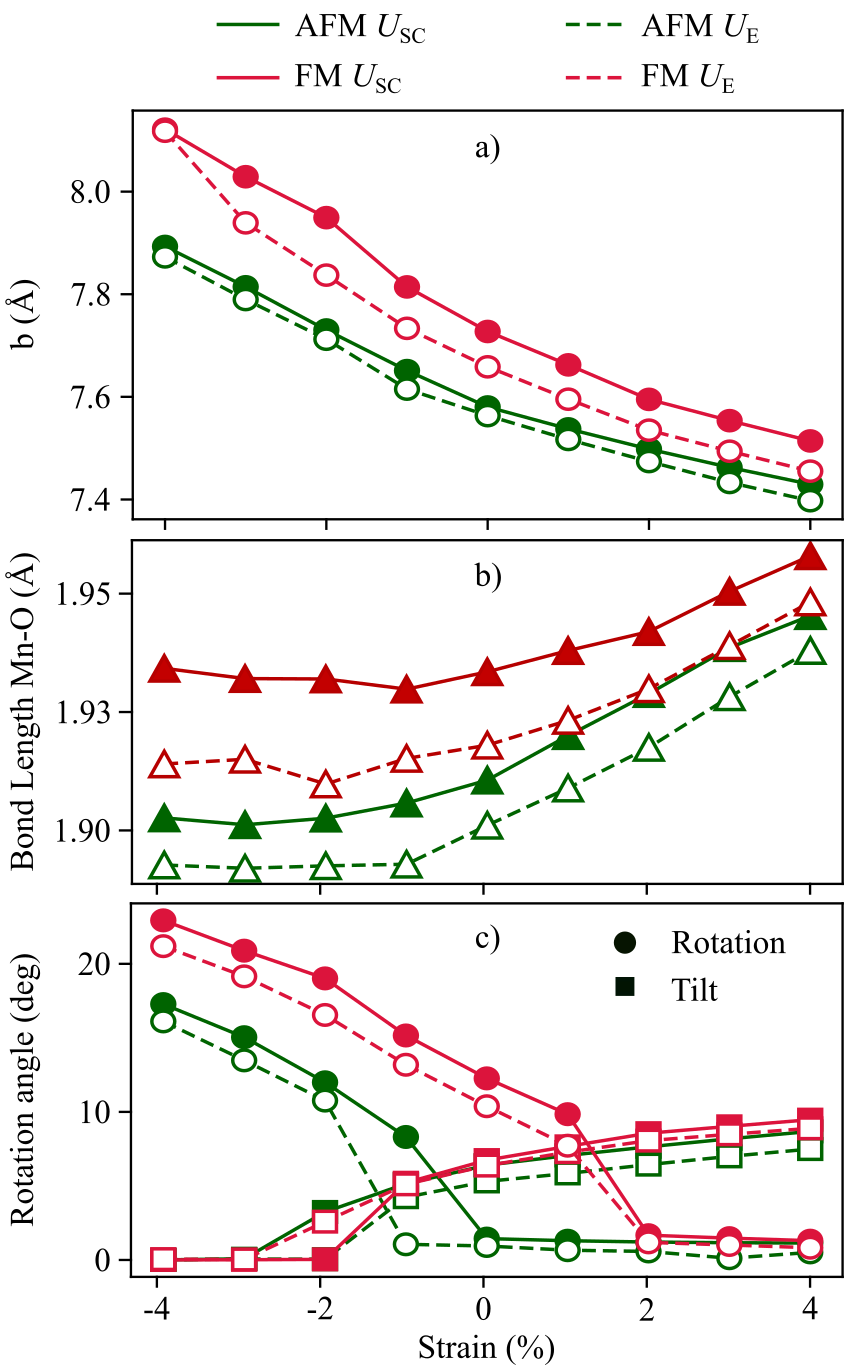}
 \caption{Dependence of the SMO structural properties on strain computed using $U_\mathrm{E}$ and $U_\mathrm{SC}$: a) lattice parameter \textit{b}, b) average Mn-O bond length, and c) octahedral tilts and rotation defined as in Fig.~\ref{fig:U_vs_structure}~(c).}
\label{fig:structural_changes_strain}
\end{figure}

We now analyze strain-induced changes of lattice parameters, namely, of Mn-O bond lengths and octahedral tilts and rotations computed with different $U$'s shown in Fig.~\ref{fig:structural_changes_strain}. As  expected on the basis of the previous discussion, $U_\mathrm{E}$ provides consistently smaller \textit{b} values, with larger differences with respect to $U_\mathrm{SC}$ for the FM phase. Mn-O bond lengths are nearly unaffected by biaxial strain, especially in the compressive range. This suggests that no significant changes in the local screening with respect to the unstrained bulk occur, in line with the small differences in $U$ computed for strained systems as discussed in the main text. This also indicates that changes in lattice parameters are mainly accommodated by changes in the octahedral rotations, which are consistently larger when using $U_\mathrm{SC}$, specially for the FM order (see Fig.~\ref{fig:structural_changes_strain}c). 

\clearpage

\section{Defective SMO}

\subsection{\label{sec:defect_types}Types of oxygen vacancies}

When a neutral oxygen vacancy is formed, we expect the two electrons, formerly associated to the removed O$^{2-}$ anion, to be accommodated on the two nearest-neighbor Mn$_1$ atoms, reducing them and changing their oxidation state from Mn$^{4+}$ to Mn$^{3+}$. In Kr{\"o}ger-Vink notation \citeSI{KROGER1956307SI}, this process is expressed as:
\begin{equation}
2 \, \textrm{Mn}^{\textrm{X}}_{\textrm{Mn}_1}+\textrm{O}^{\textrm{X}}_{\textrm{O}} \rightarrow 
2 \, \textrm{Mn}'_{\textrm{Mn}_1} + \textrm{V}_\textrm{O}^{\bullet\bullet} + \frac{1}{2} \, \textrm{O}_2 \,,
\label{eq:KV_nVO} 
\end{equation}
In this notation, the symbol designates the species (V being a vacancy), while the subscript indicates the lattice site it is located on and the superscript defines the charge relative to that lattice site, $\textrm{X}$, $\bullet$, and ' standing for neutral, positive or negative charges respectively. The above reaction states that two Mn on Mn$_1$ sites with regular oxidation state (4+) and one oxygen on an oxygen site with the regular oxidation state (2-) are converted to two reduced Mn on Mn$_1$ sites (oxidation state 3+), an oxygen vacancy, which is doubly positively charged with respect to the lattice site, and half a O$_2$ molecule released as gas. Despite the relative charge of the vacancy, this type of defect is commonly called a ``neutral'' oxygen vacancy, since only the O atom is removed, leaving behind the ions, which results in a \textit{charge neutral cell}.

Simultaneously removing one oxygen atom and one electron results in a singly positively-charged oxygen vacancy and we may expect that only one of the Mn$_1$ atoms changes its oxidation state from Mn$^{4+}$ to Mn$^{3+}$, while the other Mn$_1$ atom remains in the Mn$^{4+}$ oxidation state:
\begin{equation}
\textrm{Mn}^{\textrm{X}}_{\textrm{Mn}_1}+\textrm{O}^{\textrm{X}}_{\textrm{O}} \rightarrow \textrm{Mn}'_{\textrm{Mn}_1} + \textrm{V}_\textrm{O}^{\bullet} + \frac{1}{2} \, \textrm{O}_2 \,,
\label{eq:KV_scVO}
\end{equation}
Clearly other types of charge localization are possible (\textit{e.g.} the extra electron could be shared among the two Mn$_1$ sites), but we will show in the following that Eq. \ref{eq:KV_scVO} correctly describes a $\textrm{V}_\textrm{O}^{\bullet}$ in SMO.

Finally, a doubly positively-charged oxygen vacancy corresponds to the case when one oxygen atom and two electrons are simultaneously removed, and we expect both adjacent Mn$_1$ atoms to maintain their Mn$^{4+}$ oxidation state:
\begin{equation}
\textrm{O}^{\textrm{X}}_{\textrm{O}} \rightarrow \textrm{V}_\textrm{O}^{\textrm{X}} + 
\frac{1}{2} \, \textrm{O}_2 \,,
\label{eq:KV_dcVO}
\end{equation}

\clearpage
\subsection{\label{sec:defect_structure}Crystal structure}

Here, we discuss the effect of using different $U$ parameters ($U_\mathrm{E}$, or $U_\mathrm{SC}$, or $U_\mathrm{SC-SD}$) on the structural properties of defective SMO. The following discussion is based on defects in the AFM state. Similar conclusions can be drawn for the FM phase, for which we generally observe slightly larger relaxations. Table \ref{tbl:relaxations} shows the changes in the distance between the Mn atoms adjacent to the vacancy (Mn$_1$) as well as the maximum atomic displacement with respect to the stoichiometric bulk SMO for O atoms in nearest-neighbor positions (NN) to the defect. Geometry distortions from the stoichiometric structure are found to increase in order of V$_\textrm{O}^{\bullet\bullet}$, V$_\textrm{O}^{\bullet}$ and V$_\textrm{O}^{\textrm{X}}$. In general, the structural description provided by $U_\mathrm{SC-SD}$ coincides with the one obtained using $U_\mathrm{SC}$.

The smallest distortions are observed for V$_\textrm{O}^{\bullet\bullet}$, since the presence of the two electrons left in the lattice prevents pronounced structural changes. As we can see in Table \ref{tbl:relaxations}, atomic displacements occur predominantly along the \textit{a} direction for the NN O atoms, which move toward the O vacancy, while the position of Mn atoms adjacent to the defect remains unchanged. As a result, there is an increase in the tilt angles in the octahedral fragments adjacent to the O vacancy, which is particularly pronounced at the GGA level (about 3\textdegree). Relaxations of the NN O atoms increase with the applied $U$ together with the degree of localization of the two electrons in the defect state (see Sec. \ref{sec:defect_elprop}.)

Slightly larger relaxations are observed for V$_\textrm{O}^{\bullet}$, where one electron is removed upon the defect formation: since now only one electron is left in the lattice, the two Mn adjacent to the defect move away from each other and their distance increases by 0.15, 0.10, and 0.09 \AA\ at the GGA, GGA+$U_\mathrm{E}$, and GGA+$U_\mathrm{SC-SD}$, respectively; the NN O atoms move toward the defect by 0.12 and about 0.2 \AA\ when using GGA and GGA+$U$ methods.

The strongest structural changes are found for V$_\textrm{O}^\textrm{X}$. At the GGA and GGA+$U_\mathrm{E}$ levels, the position of both Mn$_1$ ions changes strongly: the two atoms move symmetrically away from the vacancy along \textit{a} resulting in an increase of the Mn$_1$--Mn$_1$ distance with respect to the one found in stoichiometric SMO by 0.29 or 0.28 \AA\ using GGA or GGA+$U_\mathrm{E}$, respectively. Similar distortions along \textit{a} are observed for the NN O atoms (displacements toward the vacancy of 0.14 and 0.19 \AA\ with GGA and GGA+$U_\mathrm{E}$, respectively). The GGA+$U_\mathrm{SC}$ or the GGA+$U_\mathrm{SC-SD}$ cases provide a different picture: as before, no longer feeling the repulsion of the two extra electrons as in the case of the V$_\textrm{O}^{\bullet\bullet}$ defect, the nearest O atoms undergo important changes (0.24 \AA) along \textit{a}, while the Mn position is not modified upon vacancy creation. This results in a Mn-Mn distance increased by only 0.09 \AA\ when compared to the stoichiometric bulk. The differences in results obtained with empirical or self-consistent $U$ can be explained in terms of the different electronic structure predicted with these two Hubbard parameters as we will further discuss in the following section.  

\begin{table}[h]
\caption{Changes in Mn-Mn distance between sites adjacent to a defect ($\Delta d$(Mn$_1$-Mn$_1$), in \AA) and relaxations toward the vacancy along the $a$-axis ($r_a$(NN O), in  \AA) for the O atoms in nearest-neighbors position to the defect. Results for a  V$^{\bullet \bullet}_\mathrm{O}$, V$^{\bullet }_\mathrm{O}$, V$^{\mathrm{X}}_\mathrm{O}$ in G-AFM SMO obtained with different functionals are compared. }
\small
\begin{tabular}{llcc}
\hline
\hline
	Defect		& Method			 & $\Delta d$(Mn$_1$-Mn$_1$) & $r_a$(NN O) \\
\hline
\multirow{4}{*}{V$^{\bullet \bullet}_\mathrm{O}$} & GGA 				 	&+0.05 & 0.11\\
                                         & GGA+$U_\mathrm{E}$  	&+0.00 & 0.14\\
                                         & GGA+$U_\mathrm{SC}$ 	&+0.00 & 0.15\\
                                         & GGA+$U_\mathrm{SC-SD}$	&+0.00 & 0.15\\
\hline
\multirow{4}{*}{V$^{\bullet}_\mathrm{O}$} 		 & GGA 				 	&+0.15 & 0.15\\
                                         & GGA+$U_\mathrm{E}$  	&+0.10 & 0.19\\
                                         & GGA+$U_\mathrm{SC}$ 	&+0.09 & 0.18\\
                                         & GGA+$U_\mathrm{SC-SD}$	&+0.09 & 0.18\\
\hline
\multirow{4}{*}{V$^{\mathrm{X}}_\mathrm{O}$} 				 & GGA 				 	&+0.29 & 0.14\\
                                         & GGA+$U_\mathrm{E}$  	&+0.28 & 0.19\\
                                         & GGA+$U_\mathrm{SC}$ 	&+0.09 & 0.24\\
                                         & GGA+$U_\mathrm{SC-SD}$	&+0.09 & 0.24\\
\hline
\hline
\end{tabular}
\label{tbl:relaxations}
\end {table}

\clearpage
\subsection{\label{sec:defect_elprop} Electronic Properties of Defective SMO}

\begin{figure}[h]
 \centering
 \includegraphics{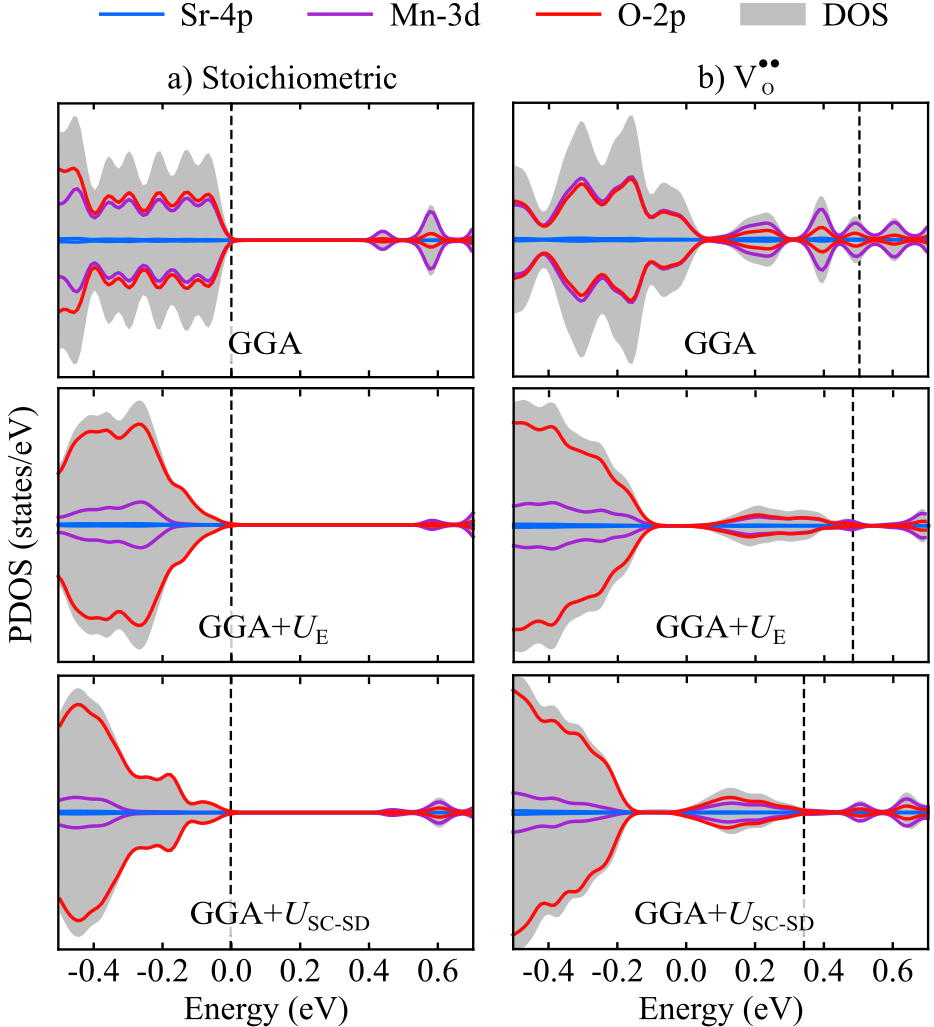}
 \caption{Total and projected density of states (DOS and PDOS, respectively) for a) the stoichiometric SMO bulk and for b) a V$_\mathrm{O}^{\bullet \bullet}$ the AFM phase of SMO computed with different functionals. The zero of energy was set at the top of the VB of the stoichiometric bulk. The vertical dotted line in b) indicates the position of the Fermi level in the defective systems.}
\label{fig:PDOS_SMO_nVO}
\end{figure}

In Figure \ref{fig:PDOS_SMO_nVO}, we compare the DOS of stoichiometric SMO with the one containing a V$_\mathrm{O}^{\bullet\bullet}$, computed using GGA, GGA+$U_\textrm{E}$ and GGA+$U_\textrm{SC-SD}$. We observe significant changes of the character of states forming the VB in the stoichiometric materials, as the Mn states are pushed towards lower energies with increasing $U$. While the materials becomes metallic upon defect formation at the GGA level, the defect state becomes increasingly more localized at the GGA+$U_\textrm{E}$ and GGA+$U_\textrm{SC-SD}$ level.  

\begin{figure}[h]
  \centering
  \includegraphics{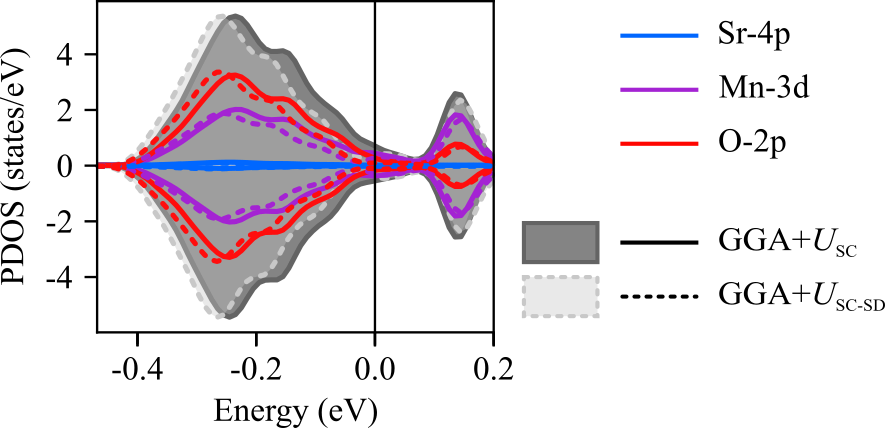}
  \caption{Zoom on DOS and PDOS of a V$_\textrm{O}^{\bullet \bullet}$ in SrMnO$_3$ in the energy interval -0.5 -- 0.2 eV using $U_\mathrm{SC-SD}$ and $U_\mathrm{SC}$. Compared to Fig.~\ref{fig:PDOS_SMO_nVO}, the zero of the energy scale is here set at the Fermi level of the defective systems.}
\label{fig:PDOS_SMO_nVO_inset}
\end{figure}

To complement the above information, we show in Fig.~\ref{fig:PDOS_SMO_nVO_inset} a zoom on the defect state, comparing results obtained for $U_\mathrm{SC}$ and $U_\mathrm{SC-SD}$. Small changes in the electronic structure are observed: the occupied defect level and the states at the bottom of the CB slightly shift toward lower or higher energies, respectively, when $U_\mathrm{SC-SD}$ is used. Even smaller changes are obtained for charged vacancies (not shown), in line with the smaller deviations of the $U_\mathrm{SC-SD}$ from $U_\mathrm{SC}$ observed in these cases. This result together with the discussion made in the previous section suggests that taking into account the site-dependence of $U$ does not significantly alter the description of the structural and electronic properties already provided by $U_\mathrm{SC}$.

\clearpage

\begin{figure}[h]
  \centering
  \includegraphics{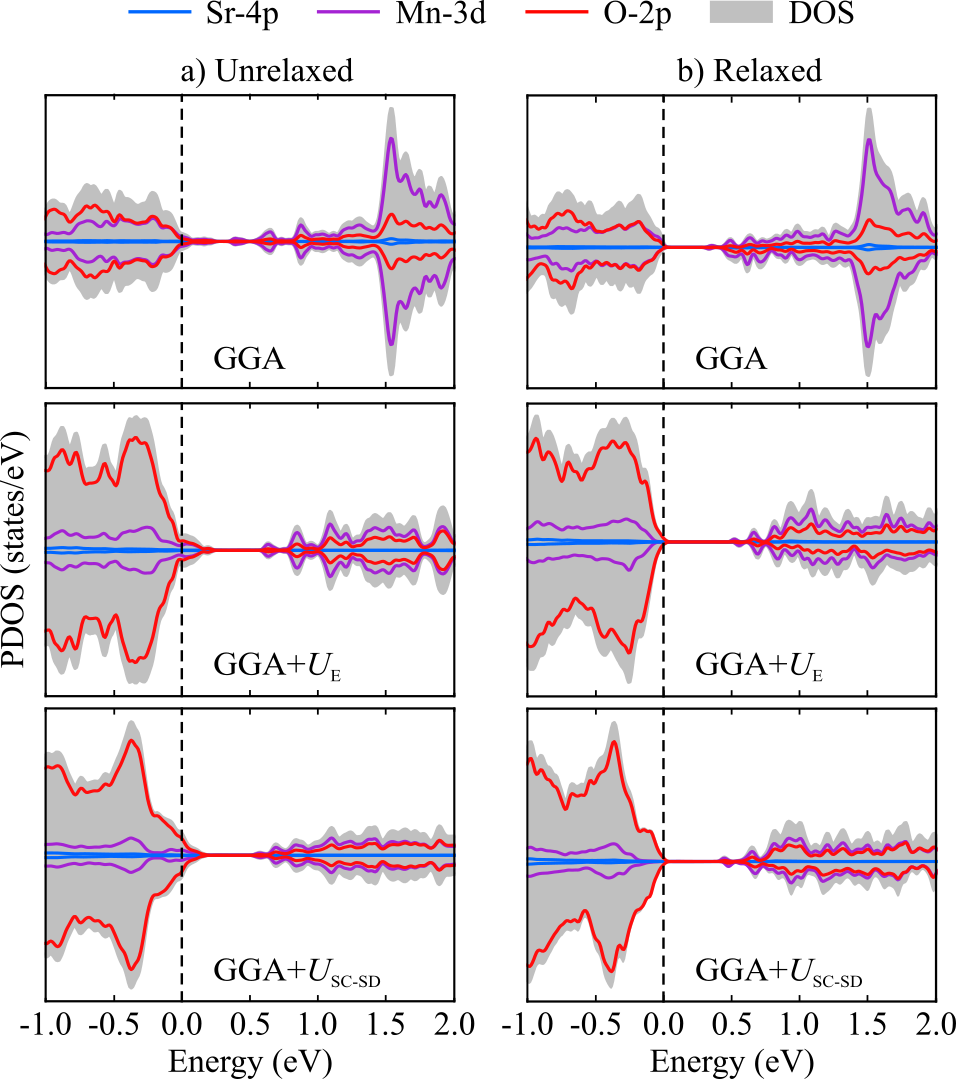}
  \caption{Projected density of states (PDOS) for a V$_\mathrm{O}^{\bullet}$ in SMO computed at the GGA, GGA+$U_\mathrm{E}$, and GGA+$U_\mathrm{SC-SD}$ level prior to a) and after b) structural optimization. The zero of the energy scale is set at the Fermi level of the defective system.}
\label{fig:PDOS_SMO_scVO}
\end{figure}

In case of a V$_\textrm{O}^{\bullet}$, one oxygen atom and one electron are removed from the supercell resulting in the reduction of only one of the Mn atoms adjacent to the defect (see Table \ref{tbl:SIos} and the discussion Sec. \ref{sec:defect_types}). The DOS of an unrelaxed V$_\textrm{O}^{\bullet}$ in SMO (Fig.~\ref{fig:PDOS_SMO_scVO}a)  shows an empty state merged with the VB of the oxide in all cases; however, the relaxations of the atoms in NN positions to the vacancy discussed in Sec. \ref{sec:defect_structure} shift this level to the bottom of the CB as shown in Fig.~\ref{fig:PDOS_SMO_scVO}b.

\begin{figure}[h]
  \centering
  \includegraphics{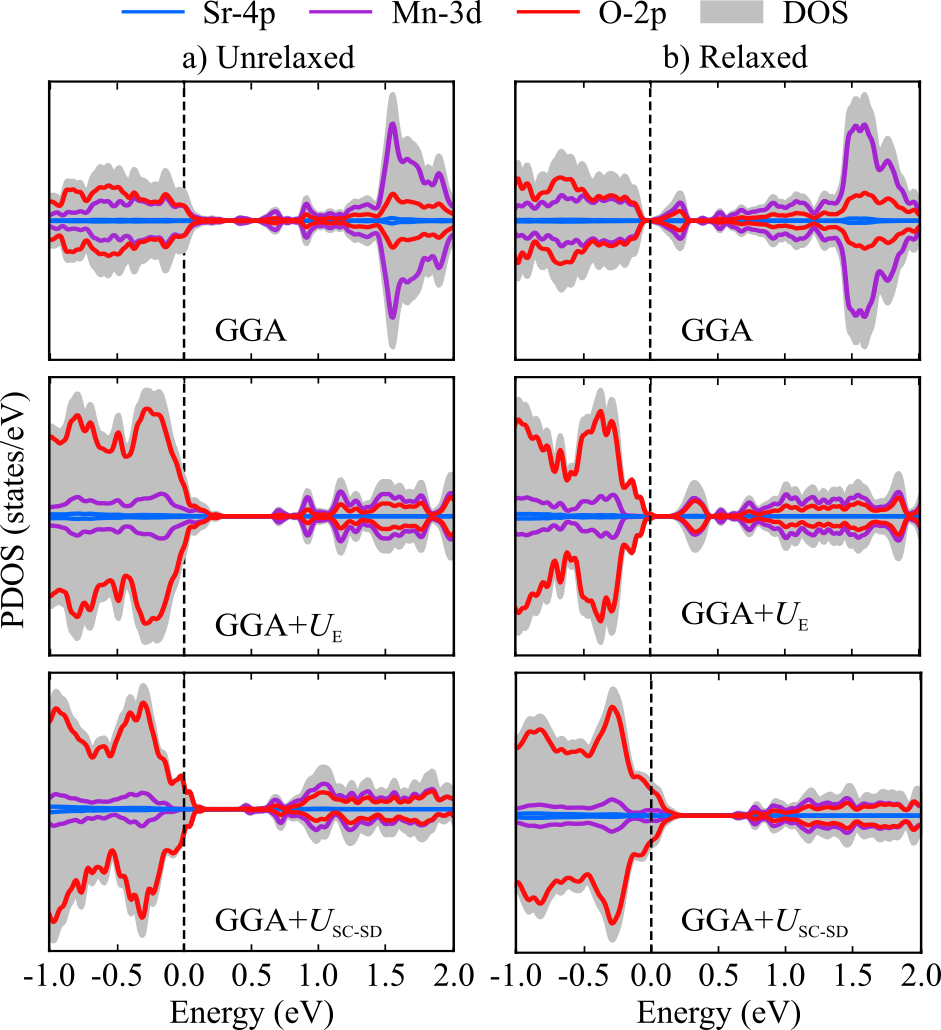}
  \caption{Projected density of states (PDOS) for  V$_\mathrm{O}^\mathrm{X}$ in SMO computed at the GGA, GGA+$U_\mathrm{E}$, and GGA+$U_\mathrm{SC-SD}$ level, prior to a) and after b)  structural optimization. The zero of the energy scale is set at the Fermi level of the defective system.}
\label{fig:PDOS_SMO_dcVO}
\end{figure}

For a V$_\textrm{O}^{\textrm{X}}$, where the removal of two electrons together with the O atom results in the adjacent Mn atoms to preserve the oxidation state of the stoichiometric material (see Table \ref{tbl:SIos}), we see, prior to structural relaxation, an empty state that appears merged with the top of the VB (Fig.~\ref{fig:PDOS_SMO_dcVO}a). Larger relaxations at the GGA and GGA+$U_\mathrm{E}$ levels (see Sec. \ref{sec:defect_structure}) push this state to higher energies, so that it is now clearly visible in the band gap, while in line with the smaller relaxations observed when $U_\mathrm{SC-SD}$ (or $U_\mathrm{SC}$) is used, the state is still merged with the VB (Fig.~\ref{fig:PDOS_SMO_dcVO}b). We can also observe how the character of this defect states changes, going from dominated by Mn-$3d$ states at the GGA level to dominated by O states for $U_\mathrm{SC-SD}$, resulting in the observed small relaxations of the neighboring Mn atoms.

\begin{table}[h]
\caption{Oxidation state (OS) and spin-up and spin-down $d$-orbital occupation numbers (Tr[$d$])  averaged over all of the Mn atoms close to the defect (Mn$_1$) and the remaining Mn sites computed in the case of V$^{\bullet \bullet}_\mathrm{O}$, V$^{\bullet }_\mathrm{O}$, V$^{\mathrm{X}}_\mathrm{O}$ in G-AFM SMO obtained at the GGA, GGA+$U_\mathrm{E}$, GGA+$U_\mathrm{SC-SD}$ level. For V$^{\bullet }_\mathrm{O}$, the two Mn$_1$ atoms show different behaviors and are hence reported separately (Mn$_1$(a) and Mn$_1$(b)). Results averaged over all Mn are also reported in the case of the stoichiometric material.}
\small
\resizebox{\linewidth}{!}{%
\begin{tabular}{llcccccccccccccccccccccc}
\hline
\hline
 & &  &\multicolumn{6}{c}{GGA}		&& \multicolumn{6}{c}{GGA+$U_\mathrm{E}$}	&& \multicolumn{6}{c}{GGA+$U_\mathrm{SC-SD}$}\\\cline{4-9}\cline{11-16}\cline{18-23}
	System		& Site			 & Spin & \multicolumn{5}{c}{Tr[d]} & OS &\multicolumn{5}{c}{Tr[d]} & OS & \multicolumn{5}{c}{Tr[d]} & OS \\
\hline
\multirow{ 2}{*}{Stoic.}& $\overline{\mathrm{Mn}}$  	& majority 	&0.658  &0.659  &0.996  &0.996  &0.96	& 4+	&&0.622  &0.623  &0.993  &0.994  &0.994 & 4+	&&0.655  &0.656  &0.996  &0.996  &0.996	& 4+\\
										 & 	  			       & minority 	&0.220 &0.223  &0.225  &0.418  &0.421	& 	&&0.262  &0.264  &0.265  &0.433  &0.435	&	&&0.223  &0.226 &0.228  &0.420  &0.422	&\\
\hline
\multirow{ 2}{*}{V$^{\bullet \bullet}_\mathrm{O}$}& Mn$_1$ 		& majority 	&0.554  &0.837  &0.979  &0.979  &0.980	& 4+	&&0.586  &0.928  &0.993  &0.993  &0.994 & 3+	&&0.606  &0.944  &0.995  &0.995  &0.995	& 3+\\
										 & 	  			& minority 	&0.263 &0.263  &0.284  &0.417  &0.424	& 	&&0.186  &0.187  &0.212  &0.353  &0.388	&	&&0.163  &0.165  &0.189  &0.330  &0.376	&\\
										 & $\overline{\mathrm{Mn}}$ 	& majority  	&0.568 &0.577  &0.983  &0.984  &0.987	& 4+	&&0.617  &0.621  &0.994  &0.994  &0.995	& 4+	&&0.651  &0.656  &0.996  &0.996  &0.996	& 4+\\
										 &              & minority   &0.328 &0.328  &0.331  &0.453  &0.455	& 	&&0.259  &0.264  &0.268  &0.434  &0.441	&	&&0.221  &0.226  &0.229  &0.420  &0.428	&\\
\hline
\multirow{ 2}{*}{V$^{\bullet}_\mathrm{O}$}		 & Mn$_1$(a)	& majority 	&0.565  &0.900  &0.977  &0.977  &0.981	& 3+	&&0.607  &0.953  &0.993  &0.993  &0.994 & 3+	&&0.640  &0.966  &0.995  &0.996  &0.996	& 3+\\
										 & 	  			& minority 	&0.264  &0.265  &0.283  &0.290  &0.425	& 	&&0.186  &0.187  &0.217  &0.226  &0.398 &	&&0.153  &0.154  &0.189  &0.194  &0.383	&\\
										 & Mn$_1$(b)	& majority 	&0.325  &0.337  &0.338  &0.461  &0.525	& 4+	&&0.258  &0.273  &0.275  &0.444  &0.488	& 4+	&&0.220  &0.236  &0.238  &0.429  &0.461	&4+\\

										 & 	  			& minority 	&0.537  &0.570  &0.979  &0.979  &0.982	& 	&&0.604  &0.609  &0.992  &0.992  &0.994 & 	&&0.638  &0.656  &0.994  &0.994  &0.996	& \\
										 & $\overline{\mathrm{Mn}}$ 	& majority  	&0.564  &0.574  &0.983  &0.984  &0.986	& 4+	&&0.616  &0.618  &0.993  &0.994  &0.995	& 4+	&&0.650  &0.652  &0.995  &0.996  &0.996	& 4+\\
										 &              & minority   &0.326  &0.330  &0.336  &0.448  &0.452	& 	&&0.261  &0.264  &0.268  &0.427  &0.437	&	&&0.220  &0.236  &0.238  &0.429  &0.461	&\\
\hline
\multirow{ 2}{*}{V$^{\mathrm{X}}_\mathrm{O}$}			& Mn$_1$ 			& majority 	&0.579  &0.597  &0.966  &0.966  &0.985	& 4+	&&0.631  &0.642  &0.990  &0.991  &0.995	& 4+	&&0.635  &0.651  &0.991  &0.992  &0.995	& 4+\\
										 & 	  			& minority	&0.332  &0.345  &0.345  &0.408  &0.468	& 	&&0.263  &0.265  &0.274  &0.370  &0.451	&	&&0.220  &0.238  &0.242  &0.427  &0.464	&\\
										 & $\overline{\mathrm{Mn}}$ 	& majority  	&0.573  &0.592  &0.976  &0.980  &0.985	& 4+	&&0.609  &0.620  &0.992  &0.994  &0.995	& 4+	&&0.641  &0.643  &0.994  &0.996  &0.996	& 4+\\
										 &              & minority   &0.311  &0.324  &0.332  &0.430  &0.449	& 	&&0.248  &0.263  &0.272  &0.423  &0.436	&	&&0.229  &0.233  &0.235  &0.413  &0.425	&\\
\hline
\hline
\end{tabular}}
\label{tbl:SIos}
\end {table}

\clearpage
\bibliographystyleSI{apsrev4-1}
\bibliographySI{references}

\end{document}